\documentclass[fleqn,10pt]{wlscirep}

\title{Optimizing the robustness of electrical power systems against cascading failures}

\author[*]{Yingrui Zhang}
\author[+]{Osman Ya\u{g}an}
\affil[]{Department of ECE, Carnegie Mellon University, Pittsburgh, 15213, USA}

\affil[*]{yingruiz@andrew.cmu.edu}
\affil[+]{oyagan@ece.cmu.edu}

\graphicspath{ {images/} }
\usepackage{graphicx}
\usepackage{subcaption}
\usepackage{bbm}
\newcommand{\bP}[1]{{\mathbb{P}}\left[{#1}\right]}
\newcommand{\bE}[1]{{\mathbb{E}}\left[{#1}\right]}

\newcommand{\1}[1]{{\bf 1}\left[#1\right]}
\newcommand{\fsquare}{\vrule height6pt width7pt depth1pt}   
\newcommand{\myendpf}{\hfill\fsquare \\[0.1in]}

\newtheorem{claim}{Claim}

\begin{abstract}
Electrical power systems are one of the most important infrastructures that support our society. However, their vulnerabilities have raised great concern recently due to several large-scale blackouts around the world. In this paper, we investigate the robustness of power systems against cascading failures initiated by a {\em random} attack. This is done under a simple yet useful model based on {\em global} and {\em equal} redistribution of load upon failures. We provide a complete understanding of system robustness by i) deriving an expression for the final system size as a function of the size of initial attacks; ii) deriving the critical attack size after which system breaks down completely; iii) showing that complete system breakdown takes place through a first-order (i.e., discontinuous) transition in terms of the attack size; and iv) establishing the optimal load-capacity distribution that maximizes robustness. In particular, we show that robustness is maximized when the difference between the capacity and initial load is the same for all lines; i.e., when all lines have the same redundant space regardless of their initial load. This is in contrast with the intuitive and commonly used setting where capacity of a line is a fixed factor of its initial load. 

\end{abstract}
\begin{document}

\flushbottom
\maketitle

\thispagestyle{empty}

\section{Introduction}
Electrical power systems are one of the most critical national infrastructures affecting all areas of daily life \cite{rinaldi2001identifying,brummitt2012suppressing}. They also provide crucial support for other national infrastructures such as telecommunications, transportation, water supply systems and emergency services \cite{o2007critical,dobson2007complex,yaugan2012optimal}. Besides daily life, the destruction of power systems would also weaken or even disable our defense and economic security \cite{foundations1997protecting}. Thus, ensuring the robustness of electrical power systems and maintaining the continuous availability of power supply are of utmost priority. Despite its great importance, concerns about the robustness of the power grid have grown recently because of several large-scale outages that took place in different parts of the world. For example, in the 2012 India blackout, 600 million people, nearly a tenth of the world's population, were left without power \cite{romero2012blackouts}. The blackout spread across 22 states in Northern, Eastern, and Northeast India and is the largest power outage in history \cite{tang2012analysis}. 

These blackouts often start with natural hazards such as lightning shorting a line or with malicious attacks, and affect only a small portion of the power system initially. But due to the long range nature of electricity, the redistribution of power loads may affect not only geographically co-located lines but also other parts of the system far from the initial affected area. A typical example is the Western Systems Coordinating Council (WSCC) system outage on August 10, 1996 \cite{kosterev1999model}, where long range failures have been observed. The large-scale blackouts are often attributed to this initial shock getting escalated due to the intricate dependencies within a power system. For example, 
when a line is tripped, the flow on all other lines will be updated, and some lines may end up with a total flow (initial plus redistributed after failures) exceeding their capacity. All lines with flows exceeding their capacity will in turn fail and flows on other lines will be updated again, possibly leading to further failures, and so on. This process may continue recursively and lead to a {\em cascade} of failures, which may potentially breakdown the entire system. For instance, on August 10, 1996, an electrical line sagged in summer heat in Southern Oregon and this initiated a chain reaction that cut power to more than four million people in eleven Western States \cite{carreras2001evidence,sachtjen2000disturbances}.

Since electrical power systems are among the largest and most complex technological systems ever developed \cite{anghel2007stochastic}, it is often hard to have a full understanding of their inter- and intra-dependencies and therefore it is hard to predict their behavior under external attacks or random failures. In this work, we aim to shed light on the robustness of power systems using a simple yet useful model. In particular, we assume that when a line fails, its load (i.e., flow) is redistributed {\em equally} among all other lines. The equal load redistribution model has the ability to capture the {\em long-range} nature of the Kirchhoff's law, at least in the mean-field sense, as opposed to the {\em topological} models where failed load is  redistributed only {\em locally} among neighboring lines \cite{CrucittiLatora,WangRong}.
This is particularly why 
this model received recent attention in the context of power systems first in the work by Pahwa et al. \cite{pahwa2010topological} and then in Ya\u{g}an \cite{yaugan2015robustness};  
the model is originally inspired by the {\em democratic fiber-bundle model} \cite{daniels1945statistical} that is used extensively for studying the rupture of fiber-bundles under increasing external force.

Assuming that each of the $N$ lines in the system is assigned independently an initial load $L_i$ and a redundant space $S_i$ -- meaning that line capacity equals $L_i+S_i$ -- from a joint distribution $P_{LS}(x,y)= \bP{L \leq x, S \leq x}$, we study the robustness of this systems against {\em random} attacks; see Section \ref{sec:Model} for the details of our model. In particular, we characterize the {\em final} fraction $n_{\infty}(p)$ of alive (i.e., non-failed) lines at the steady-state, when $p$-fraction of the lines are randomly failed. We identify the critical attack size $p^{\star}$ after which the system breakdowns entirely; i.e., $n_{\infty}(p)=0$ if $p > p^{\star}$. We show that the transition of the system around $p^{\star}$ is always first-order (i.e., discontinuous). However, depending on the distribution $P_{LS}$, this may take place with or without a preceding second-order (i.e., continuous) divergence of $n_{\infty}(p)$ from the $1-p$ line. Finally, under the constraints that mean load $\bE{L}$ and mean free space $\bE{S}$ are fixed, we show that assigning every line the same free space regardless of its load is {\em optimal} in the sense of maximizing the robustness $n_{\infty}(p)$ for all attack sizes $p$. This provably optimal strategy is in sharp contrast with the commonly used
\cite{MotterLai,WangChen,Mirzasoleiman,CrucittiLatora,bernstein2014power,kinney2005modeling} setting where free-space $S_i$ is set to be a constant factor of a line's initial load, e.g.,  $S_i = \alpha L_i$ for all $i$. This hints at the possibility that existing power systems are not designed optimally and that their robustness may be significantly improved by reallocating the line capacities (while keeping the total capacity unchanged).
Our analytic results are validated via extensive simulations, using both synthetic data for load-capacity values as well as realistic data from IEEE test-case data sets.

We believe that our results provide interesting insights into the dynamics of cascading failures in power systems. In particular, we expect our work to shed some light on the {\em qualitative}
behavior of real-world power systems under random attacks, and help 
 design them in a more robust manner. The results obtained here may have applications in fields other than power systems as well. A particularly interesting application is the study of the traffic jams in roads, where the capacity of a line can be regarded as the traffic flow capacity of a road \cite{su2014robustness,scala2015cascades}.

\section{Model Definitions}
\label{sec:Model}
{\bf Equal load-redistribution model.} We consider a power system with $N$ transmission lines $\mathcal{L}_1, \ldots, \mathcal{L}_N$ 
with initial loads (i.e., power flows) $L_1, \ldots, L_N$.
The {\em capacity} $C_i$  of a line $\mathcal{L}_i$ defines the maximum power flow that it can sustain, and 
is  given by 
\begin{equation}
C_i= L_i + S_i, \qquad i=1,\ldots, N,
\label{eq:capacity}
\end{equation}
where $S_i$ denotes the {\em free-space} (or, redundancy) available to line $\mathcal{L}_i$. 
The capacity of a line can be defined as a factor of its initial load, i.e.,
\begin{equation}
C_i = (1+\alpha_i) L_i
\label{eq:capacity_with_tolerance}
\end{equation}
with $\alpha_i>0$ defining the {\em tolerance} parameter for line $\mathcal{L}_i$. Put differently, the free space $S_i$ is given in terms of
the initial load $L_i$ as $S_i = \alpha_i L_i$; 
it is very common \cite{MotterLai,WangChen,Mirzasoleiman,CrucittiLatora} to use a {\em fixed} tolerance factor for all lines in the system, i.e.,  to use $\alpha_i=\alpha$ for all $i$.
It is assumed that a line {\em fails} (i.e., outages) if its load exceeds its capacity at any given time. The key assumption of our model is that when a line fails, the load it was carrying (right before the failure)
is redistributed {\em equally} among all remaining lines. 

Throughout we assume that the pairs  $(L_i, S_i)$ are independently and identically distributed with $P_{LS}(x,y):=\bP{L \leq x, S \leq y}$ for each $i=1,\ldots, N$. The corresponding (joint) probability density
function is given by $p_{LS}(x,y) = \frac{\partial^2}{\partial x \partial y} P_{LS}(x,y)$.
Throughout, we let $L_{{\sl min}}$ and $S_{{\sl min}}$ denote the minimum values for load $L$ and free space $S$; i.e.,
\begin{align}
L_{{\sl min}} &= \inf\{x: P_L(x) > 0\}
\nonumber \\ \nonumber
S_{{\sl min}} &= \inf\{y: P_S(y) > 0\}
\end{align}
We assume that $L_{{\sl min}},  S_{{\sl min}} > 0$. We also assume that the marginal densities $p_L(x)$ and $p_S(y)$ are continuous on their support.

The equal load redistribution rule takes its roots from the {\em democratic} fiber bundle model \cite{andersen1997tricritical,daniels1945statistical},
where $N$ parallel fibers with failure thresholds $C_1, \ldots, C_N$ (i.e., capacities) 
share an applied total force $F$ {\em equally}. There, it has been of interest to study the dynamics of recursive failures in the bundle as the applied force
$F$ increases;
e.g., see 
\cite{da1998comment,sornette1997conditions,pradhan2003failure,roy2015fiber}. This model has been recently used by Pahwa et al. \cite{pahwa2014abruptness}  in the context  of power systems, with $F$ corresponding to the total load shared {\em equally} by $N$ power lines.
The relevance of the equal load-redistribution model for power systems stems from its ability to capture the {\em long-range} nature of the Kirchoff's law, at least in the mean-field sense, as opposed to the {\em topological} models where failed load is  redistributed only {\em locally} among neighboring lines \cite{CrucittiLatora,WangRong}. 
In particular, the equal load redistribution model is expected to be a reasonable
assumption under  the DC power flow model, which approximates the standard AC power flow model 
when the phase differences along the branches are small and the bus voltages are 
fixed \cite{pahwa2014abruptness}; in fact,  power flow calculations based on the DC model \cite{Overbye,StottJardimAlsac} are known to give accurate results that match the AC model calculations in many cases. 
Therefore, we expect our work to shed some light on the {\em qualitative}
behavior of real-world power systems under random attacks.

 {\bf Problem definition.}
Our main goal is to study the robustness of power systems under the equal load redistribution rule.
In this work, we assume that failures are initiated by a 
{\em random} attack that results with a failure of a $p$-fraction of the lines; of course, all the discussion and 
accompanying results do hold for the robustness against random {\em failures} as well.
The initial failures lead to redistribution of power flows from the failed lines to {\em alive} ones (i.e., non-failed lines), 
so that the load on each alive line becomes equal to its initial load plus its equal share of the total load of the failed lines.  This may lead to 
the failure of some additional lines due to 
the updated flow exceeding their capacity. This process may continue recursively, generating a {\em cascade of failures}, 
with each failure further increasing the load on the alive lines, and may eventually result
with the collapse of the entire system. Throughout, we let $n_{\infty}(p)$ denote the {\em final} fraction of alive lines
when a $p$-fraction of lines is randomly attacked. The {\em robustness} of a power system will be evaluated 
by the behavior $n_{\infty}(p)$ for all attack sizes $0<p<1$. Of particular interest is to characterize the {\em critical} attack size $p^{\star}$ at which 
$n_{\infty}(p)$ drops to zero.

The problem formulation considered here was introduced by Ya\u{g}an \cite{yaugan2015robustness}.
This formulation differs from the original democratic fiber-bundle model (and its analog \cite{pahwa2014abruptness} introduced for power systems)
in that i) it does not assume that  the total load of 
 the system is fixed at $F$; and ii) it allows for power lines to carry different initial loads unlike the democratic fiber bundle model where all lines start with the same initial load.
Since  
power lines in real systems are likely to have different loads at the initial set-up, we believe our formulation is more suitable for studying cascading failures in power systems. 
In addition,  \cite{pahwa2014abruptness} is concerned with failures in the power system 
 that are triggered by increasing the total force (i.e., load) applied.
Instead, our formulation allows analyzing the robustness of the system against external attacks
or random line failures, which are known to be the source of system-wide blackouts in many interdependent systems \cite{Rosato,Buldyrev,yaugan2012optimal}. 

A word on notation in use: The random variables (rvs)
under consideration are all defined on the same probability space
$(\Omega, {\cal F}, \mathbb{P})$. Probabilistic statements are
made with respect to this probability measure $\mathbb{P}$, and we
denote the corresponding expectation operator by $\mathbb{E}$.
The indicator function of an event $A$ is denoted by $\1{A}$. 

\section{Results}


\subsection{Final system size as a function the attack size, $n_{\infty}(p)$}
\label{subsec:Results_1}

Our first main result characterizes the robustness of power systems under any initial load-space distribution $P_{LS}$ and
any attack size $p$.
Let $L$ and $S$ denote generic random variables following the same distribution with initial loads $L_1,\ldots,L_N$,
and free spaces $S_1, \ldots, S_N$, respectively. Then, with $x^{\star}$ denoting the smallest solution of
\begin{equation}
h(x):=\bP{S > x} \left( x + \bE{L ~|~ S > x} \right) \geq \frac{\bE{L}}{1-p}
\label{eq:main_condition}
\end{equation}
over the range $x^{\star} \in (0, \infty)$, the final system size $n_{\infty}(p)$ at attack size $p$ is given by 
\begin{equation}
n_{\infty}(p) = (1-p) \bP{S > x^{\star}}.
\label{eq:final_size}
\end{equation}
This result, proved in Section \ref{subsec:methods_1}, provides a complete picture about a power system's robustness against random attacks of arbitrary size. 
In particular, it helps understand the response $n_{\infty} (p)$  of the system to attacks of varying magnitude.

For a graphical solution of $n_{\infty}(p)$, 
one shall plot $\bP{S > x} \left(x + \bE{L ~|~ S > x} \right)$ as a function of $x$ 
(e.g., see Figure \ref{fig:h_x_1}), and draw a horizontal line at the height $\bE{L}/(1-p)$ on the same plot. The leftmost intersection of these two lines gives the operating point $x^{\star}$, from which we can compute  $n_{\infty}(p) = (1-p)\bP{S > x^{\star}}$. When there is no intersection,  we set $x^{\star}=\infty$ and understand that $n_{\infty}(p)=0$.

We see from this result that an adversarial attack aimed at a certain part of the electrical power grid may lead to  failures in other parts of the system, possibly creating a recursive failure process also known as {\em cascading failures}. This will often  result with a damage in the system much larger than the initial attack size $p$. However, in most cases \lq\lq some" part of the system is expected to continue its functions by undertaking extra load; e.g., with $n_{\infty}(p)>0$. In such cases, although certain service areas are affected, the power grid remains partially functional. The most severe situations arise when cascading failures continue until the {\em complete} breakdown of the system where all lines fail; e.g., when $n_{\infty}(p)=0$. This prompts us to characterize the {\em critical} attack size $p^{\star}$, defined as the largest attack size that the system can sustain.


{\bf The \lq\lq critical" attack size.} Of particular interest is to derive the {\em critical} attack size $p^{\star}$ such that for any attack with size $p > p^{\star}$, the system undergoes a complete breakdown leading to 
$n_{\infty} (p) = 0$; on the other hand for $p < p^{\star}$, we have $n_{\infty}(p)>0$. More precisely, we define $p^{\star}$ as
\[
p^{\star} = \sup\{p: n_{\infty}(p) > 0 \}.
\]

The critical attack size can be derived from
the previous results (\ref{eq:main_condition})-(\ref{eq:final_size}) that characterize $n_{\infty}(p)$.
Namely, for any load-free space distribution $p_{LS}(x,y)$,
the maximum attack size $p^{\star}$ can be  computed from  the {\em global} maximum of the function $\bP{S > x} \left(x + \bE{L ~|~ S > x} \right)$. 
In particular, we have
\begin{equation}
p^{\star} = 1- \frac{\bE{L}}{\max\limits_{x}\{\bP{S > x} \left( x + \bE{L ~|~ S > x} \right)\}}.
\label{eq:max_attack}
\end{equation} 
A proof of this result is given in  Section \ref{subsec:methods_1}.


\subsection{Understanding the \lq\lq phase transition": Conditions for abrupt rupture.}
\label{subsec:Results_2}

It is of significant interest to understand the behavior of the system near the {\em phase transition}; i.e., 
when the attack size is very close to but smaller than the critical value $p^{\star}$. One main
questions here is whether $n_{\infty}(p)$ decays to zero continuously (i.e., through a second-order transition),
or discontinuously (i.e., through a first-order transition).
The practical significance of this is that continuous transitions  suggest a more stable and predictable system behavior with respect to attacks, whereas with discontinuous transitions system behavior becomes more difficult to predict, for instance, from past data.

\begin{figure*}[!tbp]
  \begin{subfigure}[b]{0.45\textwidth}
    \includegraphics[width=\textwidth]{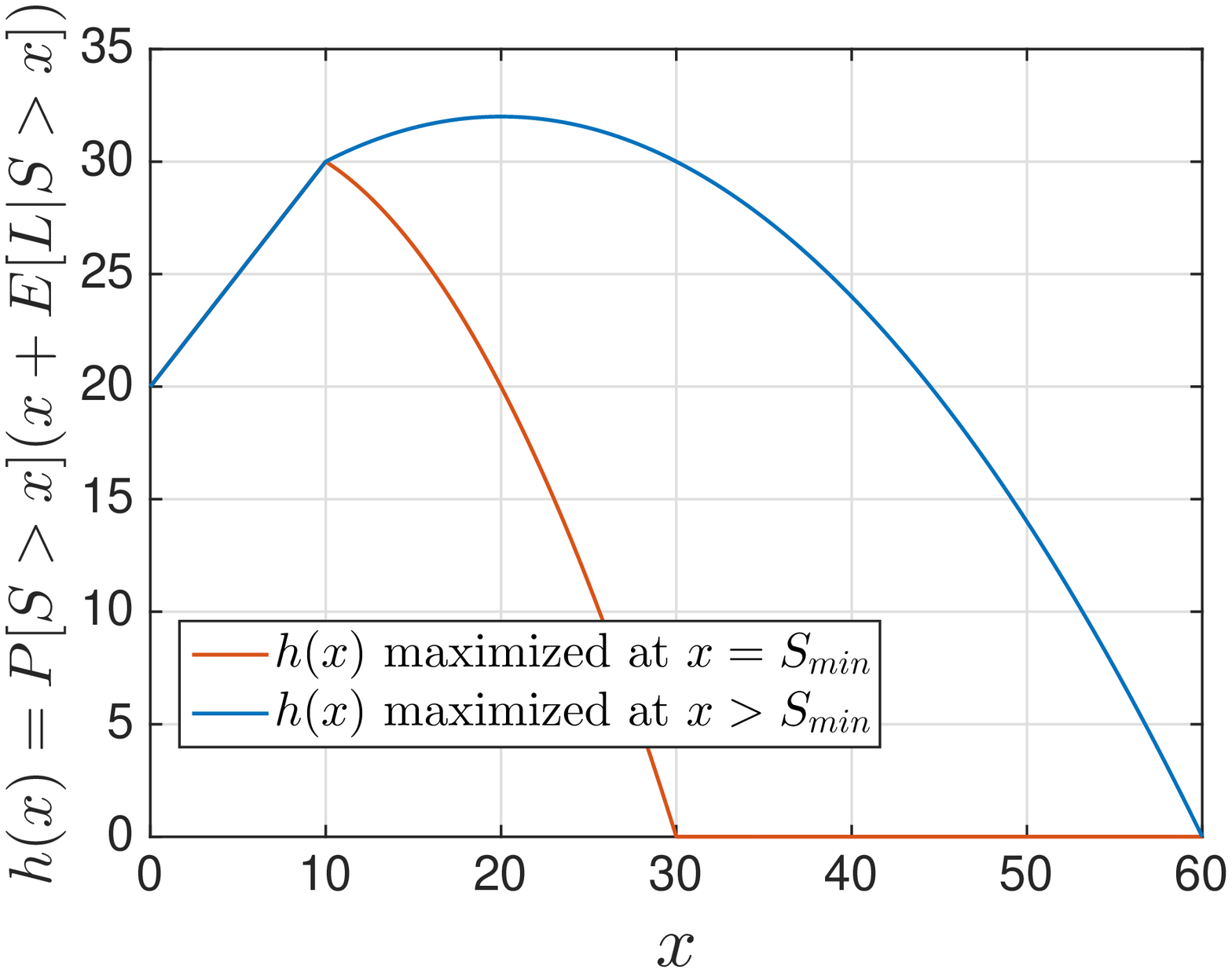}
    \caption{}
    \label{fig:h_x_1}
  \end{subfigure}
  \hfill
  \begin{subfigure}[b]{0.45\textwidth}
    \includegraphics[width=\textwidth]{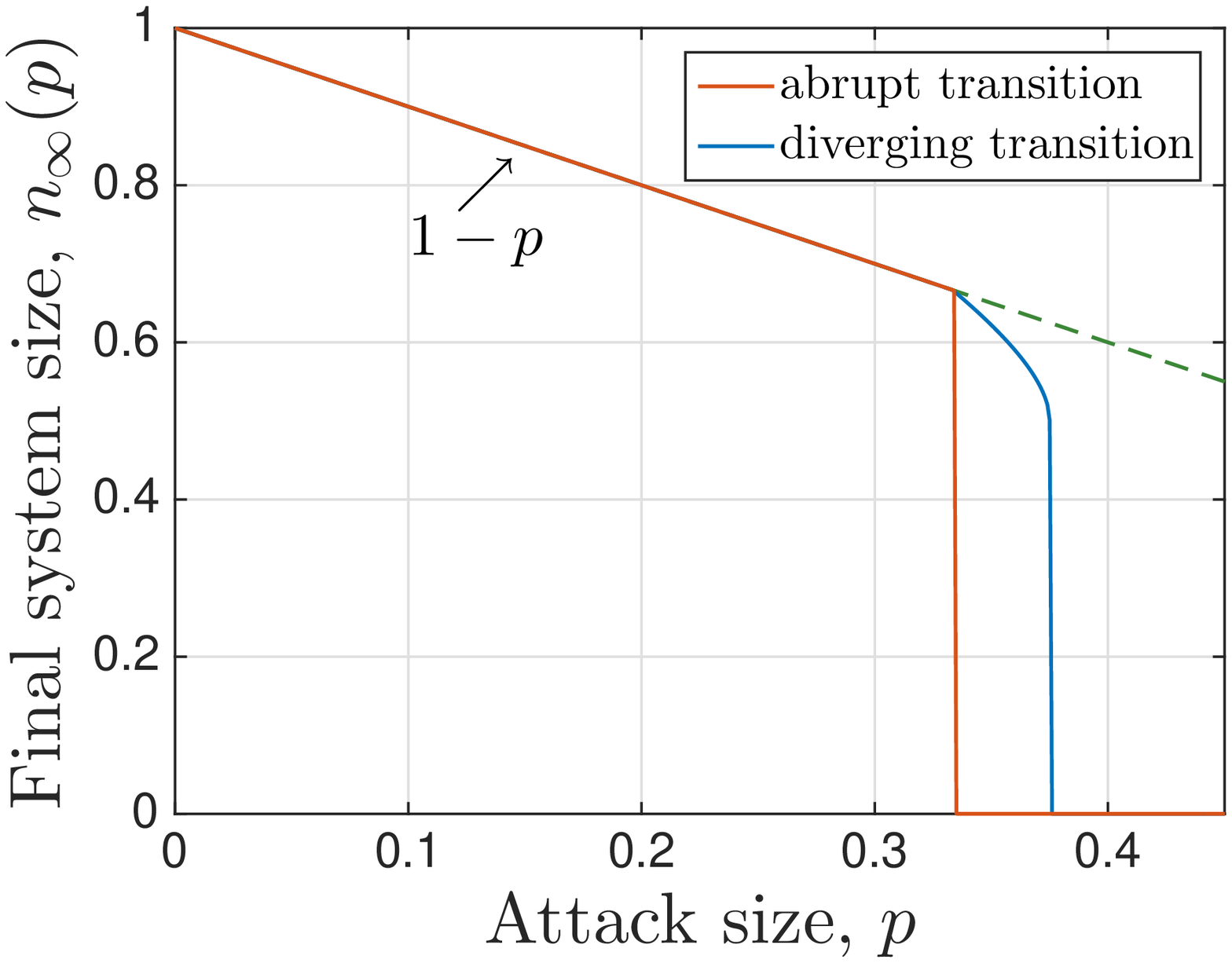}
    \caption{}
    \label{fig:h_x_2}
  \end{subfigure}
  \caption{\sl \textbf{Different types of first-order transitions.} We demonstrate the difference between an {\em abrupt} first-order transition and a first-order transition with a preceding divergence from the $1-p$ line. The lower curves (shown in red) correspond to the case where the load $L$ and extra space $S$ are independent and  uniformly distributed with $L_{min} = S_{min}= 10$ and $\mathbb{E}[L]=\mathbb{E}[S]=20$.
The upper curves (shown in blue) are obtained under the same setting except that we set $\mathbb{E}[S]=35$. We see that the lower curve in Figure \ref{fig:h_x_1} reaches its maximum at $S_{min} = 10$, and the corresponding final system size exhibits an {\em abrupt} first-order transition  as shown in Figure \ref{fig:h_x_2}. On the other hand, the upper (i.e., blue) curve   in Figure \ref{fig:h_x_1} is maximized at $S = 20 > S_{min}$. As expected from our result (e.g., see (\ref{eq:abrupt_condition})), the  total breakdown of the system takes place after a diverging failure rate is observed.}
  \label{fig:h_x}
\end{figure*}

Our analysis shows that under the equal-load redistribution model considered here the total breakdown of the system  will always be through a first-order (i.e., discontinuous) transition; see Methods for a proof. Namely, we have
\begin{align}
n_{\infty} (p^{\star}) > 0,
\label{eq:first_order}
\end{align}
while by definition it holds that
$
n_{\infty} (p^{\star}+\epsilon) = 0$, 
for any $\epsilon>0$ arbitrarily small.
This means that regardless of the attack size and the distribution of load and capacity, the transition point where the system has a total breakdown (i.e., where the fraction of alive lines drops to zero) is always discontinuous. 
These cases are reminiscent of the real-world phenomena 
of unexpected large-scale system collapses; i.e., cases where seemingly identical attacks/failures leading to entirely different consequences.

Now that we showed that the breakdown of the power system  takes place through a first-order transition,  
an interesting question arises as to whether this first-order rupture at $p^{\star}$ has any early indicators at smaller attack sizes; e.g., a 
{\em diverging} failure rate leading to a non-linear decrease in $n_{\infty} (p)$.
Otherwise, an {\em abrupt} first-order transition is said to take place if the linear decay of $n_{\infty} (p)$ (of the form $1-p$) is followed by a sudden discontinuous jump to zero at  $p^{\star}$; i.e., we say that the system exhibits an {\em abrupt} rupture when it holds that 
\begin{equation}
n_{\infty}(p) = \left\{ 
\begin{array}{cc}
1- p  & \textrm{if $p \leq p^{\star}$}     \\
 0   &      \textrm{if $p > p^{\star}$}
\end{array}
\right.
\label{eq:abrupt_rupture_defn_forme}
\end{equation}
In Figure \ref{fig:h_x_2} we demonstrate the distinction between an {\em abrupt} rupture and a rupture with preceding  divergence from the $1-p$ line.

We now present our result that reveals the necessary and sufficient condition for an abrupt rupture to take place. 
We show (in Methods) that the system goes through an abrupt first-order breakdown (e.g., see the below line shown in red in Figure
\ref{fig:h_x_2}), if and only if the function $h(x)=\mathbb{P}[S>x](x+\mathbb{E}[L \mid S>x])$ reaches its maximum at $x=S_{min}$, where $S_{min}$ is the minimum value the extra space $S$ can take. Namely, an abrupt first-order rupture ({\em without} a preceding divergence) takes place if and only if 
\begin{equation}
\arg \max_{x > 0} \left\{\bP{S > x} \left(x + \bE{L ~|~ S > x} \right) \right\} = S_{\textrm{min}}.
\label{eq:abrupt_condition}
\end{equation}
Otherwise, if $\arg \max_{x > 0} h(x) \neq S_{\textrm{min}}$, then a preceding divergence from the $1-p$ line will be observed before $n_{\infty}(p)$ drops to zero; e.g., see the above line shown in blue in Figure
\ref{fig:h_x_2}). More precisely, it will hold that $n_{\infty}(p)<1-p$ for some  $p<p^{\star}$. A detailed analysis of conditions for these two types of ruptures is presented in Methods.

Figure \ref{fig:h_x} demonstrates different types of transitions that the system can exhibit in relation to the behavior of $h(x)$. In figure \ref{fig:h_x_1}, we plot $h(x)$ in two different cases: the red (i.e., lower) line reaches its maximum at $S_{min}$, while the blue (i.e., upper) line continues to increase after $S_{min}$ and reaches its maximum later. Since the function $h(x)$ represented by the blue line does not satisfy the abrupt rupture condition (\ref{eq:abrupt_condition}), we see in figure \ref{fig:h_x_2} that the corresponding final system size goes through a diverging transition (from the $1-p$ line) before entirely breaking down through a first-order transition. 
On the other hand, we see that $h(x)$ represented by the red curve reaches its maximum at $S_{min}$. As expected from our results, we see that the corresponding final system size exhibits an abrupt breakdown without any preceding divergence from the $1-p$ line. 



\subsection{Achieving optimal robustness}
\label{subsec:Results_3}
The most important question from a system design perspective is 
concerned with deriving the {\em universally optimum} distribution of initial loads $L_1, \ldots L_N$ 
and free spaces $S_1, \ldots, S_N$ that leads to {\em maximum} robustness
under the constraints that $\bE{L}$ and $\bE{S}$ are fixed. We believe that the answer to this problem would be
very useful in designing real-world power grids with optimum robustness, i.e. with the final system size $n_{\infty}(p)$ 
maximized
for any attack size $p$. The motivation for the constraints on the mean load $\bE{L}$ and mean free space $\bE{S}$
are as follows. The total load carried by the system
is likely to be dictated by system requirements in most real-world cases, which also determines
the average load per line. In addition, the total capacity (or, total free space) available to the system is likely to be bounded due to the {\em costs} associated with using high-capacity lines. 

Our results concerning this important problem are presented next. First, we focus on maximizing the critical attack size $p^{\star}$.
We show in Methods that the critical attack size always satisfies
\begin{equation}
p^{\star} \leq \frac{\mathbb{E}[S]}{\mathbb{E}[S] + \mathbb{E}[L]} = \frac{\bE{S}}{\bE{C}}
\label{eq:max_attack_size}
\end{equation}
Namely, regardless of the distribution $p_{LS}$ that generates load-capacity pairs, the system will always go into a complete breakdown if more than
${\mathbb{E}[S]}/\mathbb{E}[C]$-fraction of lines are attacked; i.e., the system can never sustain a random attack of size exceeding the ratio of mean free space to mean capacity. Next, we show that this critical attack size is in fact attainable {\em under any load distribution} by a {\em Dirac} delta distribution for the free-spaces, i.e., by giving every line the same free space. More precisely, let $p^{\star}_{dirac}$ 
denote the  critical attack size when $p_{LS}(x,y) = p_L(x) \delta(y-\bE{S})$, where
the distribution $p_L(x)$ of the initial loads $L_1, \ldots, L_N$ is arbitrary. We show in Methods that 
\[
p^{\star}_{dirac} \geq \frac{\mathbb{E}[S]}{\mathbb{E}[S] + \mathbb{E}[L]}.
\]
Combined with (\ref{eq:max_attack_size}) this shows that assigning every line the same free space (regardless of the initial loads)
maximizes the largest attack that the system can sustain. 

More can be said regarding the optimality of equal free-space allocation. Let $p^{\star}_{optimal}$ denote the maximum critical attack size as established above, i.e., $p^{\star}_{optimal}=\mathbb{E}[S]/(\mathbb{E}[S] + \mathbb{E}[L])$. In view of the fact that we always have $n_{\infty}(p) \leq 1-p$,
the next result firmly establishes that using the Dirac delta distribution for free space optimizes the robustness of the system uniformly for
any attack size $p$. In particular, if $p_{LS}(x,y) = p_L(x) \delta(y-\bE{S})$, then the corresponding final system size $n_{\infty,dirac}(p)$ satisfies
\begin{equation}
n_{\infty,dirac}(p) = \left \{  
\begin{array}{cc}
1-p  &  \textrm{for $p < p^{\star}_{optimal} $}   \\
0  &     \textrm{for $p \geq p^{\star}_{optimal} $}
\end{array}
\right.
\label{eq:robustness_of_dirac_S}
\end{equation}
Namely, the distribution $p_{LS}(x,y) = p_L(x) \delta(y-\bE{S})$ maximizes the final system size $n_{\infty}(p)$ uniformly for all $p$.

This result shows that as far as the {\em random} attacks are concerned, the system's robustness can be maximized under the constraints of fixed $\bE{L}$ and fixed $\bE{S}$ (and hence fixed
$\bE{C}$), by
giving each line an equal free space $\bE{S}$, 
{\bf irrespective of how the initial loads are distributed}.
Put differently, the robustness will be maximized by choosing a line's capacity $C_i$ through $C_i = L_i + \bE{S}$ no matter what its load $L_i$ is. In view of (\ref{eq:capacity_with_tolerance}), this then leads to a tolerance factor 
\begin{equation}
\alpha_i =\bE{S}/L_i
\label{eq:optimal_tolerance_factor}
\end{equation}
meaning that the optimal robustness is achieved when lines with larger initial loads are given 
smaller tolerance factors. This result is rather counter-intuitive because one might think that lines with large initial loads 
shall receive extra protection (in the form of larger free space or tolerance factor) given the potentially detrimental effect of their failure to the overall system. Our result prove this intuition incorrect and show that robustness is maximized if all lines share the fixed total free-space equally. In fact, numerical results presented in Section \ref{subsec:numerical} show that the standard choice of setting free-space to be a constant factor of initial load (i.e., using the same tolerance factor for all lines) may lead to  significantly worse robustness than the optimal choice given at (\ref{eq:optimal_tolerance_factor}).

A possible explanation to this counter-intuitive result is as follows.
When all lines have the same extra space, we ensure that the system never goes through a {\em cascade} of failures. In other words, when $p$-fraction of the lines are attacked, we will have either $n_{\infty}(p)=1-p$ or $n_{\infty}(p)=0$ depending on whether or not, respectively, the total load of failed lines divided by $1-p$ is less than the common free space $S$. In addition, if the attack size is large enough that total load of failed lines, i.e., $p \bE{L}$, is larger than the total free space $(1-p) \bE{S}$ available in the rest of the system, then regardless of the distribution $p_{LS}(x,y)$, the system will collapse. Collectively, these explain why assigning equal free-space to all lines ensures that system will go through an abrupt rupture, but only at the optimal critical attack size $p^{\star}_{optimal}$.

The optimality results presented here shed light on the recent findings by Ya\u{g}an \cite{yaugan2015robustness} who investigated the model considered here in the special case where $S_i = \alpha L_i$ for all lines; i.e., the case
where $p_{LS}$ is degenerate with $p_{LS}(x,y) = p_L(x) \delta(y - x \alpha)$.
There, they found that the optimal robustness is achieved (i.e., $n_{\infty}(p)$ is maximized uniformly across all $p$), if the {\em load} $L$ follows a Dirac delta distribution; i.e., the system is most robust when each line carries the same initial load. On the other hand, our results show that the distribution of load has very little to do with optimizing the robustness, and in fact robustness is maximized under {\em any} initial load distribution if the free-space is distributed equally. This shows that Ya\u{g}an's result of the optimality of equal load distribution is merely a coincidence. It only arises under the assumption that a line's free space is a constant factor of its load, so that equal allocation of initial loads is equivalent to that of free-space.


\begin{figure*}[!t]
	\includegraphics[width=0.9\textwidth]{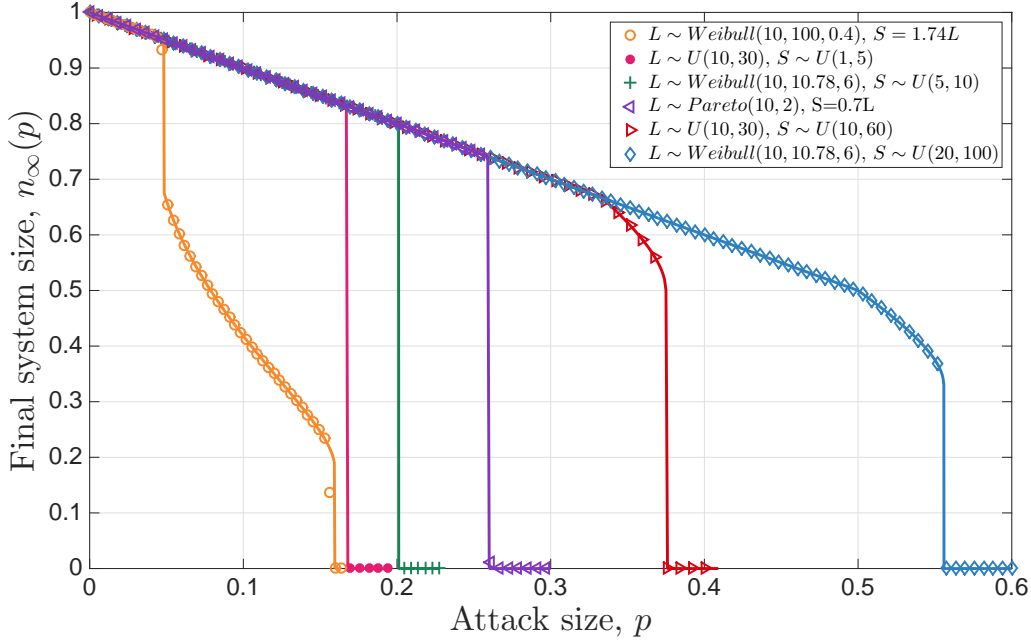}
    \centering
\caption{\sl \textbf{Final system size under different load-free space distributions.} Analytic results (obtained from (\ref{eq:main_condition}) and (\ref{eq:final_size})) are represented by lines, whereas simulation results (averaged over 200 independent runs) are represented by symbols. We see that in each case theoretical results match the simulation results very well. 
}
    \label{fig:general}
\end{figure*}

\subsection{Numerical results}
\label{subsec:numerical}
We now confirm our theoretical findings via numerical simulations, using both synthetic and real-world data. We focus on the former case first and consider various commonly known distributions for the load and free-space variables. 

\vspace{2mm}
\noindent{\bf Synthetic data.} Throughout, we consider three commonly used families of distributions: i) Uniform, ii) Pareto, and iii) Weibull. The corresponding probability density functions are defined below for a generic random variable $L$.
\begin{itemize}
\item Uniform Distribution: $L \sim U(L_{\textrm{min}},L_{\textrm{max}})$. The density is given by
\[
p_L(x)=\frac{1}{ L_{\textrm{max}}  - L_{\textrm{min}} } \cdot \1{ L_{\textrm{min}} \leq x  \leq L_{\textrm{max}} }
\]
\item Pareto Distribution: $L \sim {Pareto}(L_{\textrm{min}}, b)$. With
$L_{\textrm{min}}>0$ and $b>0$, the density is given by
\[
p_L(x) =  L_{\textrm{min}}^{b} b x^{-b-1} \1{x \geq L_{\textrm{min}}}.
\]
To ensure that $\bE{L}=b L_{\textrm{min}}/(b-1)$ is finite, we also enforce $b>1$. Distributions belonging to the Pareto  family are also known as a {\em power-law} distributions and have been extensively used in many fields including power systems.
\item Weibull Distribution: $L \sim Weibull(L_{\textrm{min}},\lambda, k)$. With $\lambda, k, L_{\textrm{min}}>0$, the density is given by
\[
p_L(x) = \frac{k}{\lambda} \left(\frac{x-L_{\textrm{min}}}{\lambda} \right)^{k-1} e^{-\left(\frac{x-L_{\textrm{min}}}{\lambda} \right)^{k}}
 \1{x \geq L_{\textrm{min}}}.
 \]
The case $k=1$ corresponds to the exponential distribution, and $k=2$ corresponds to Rayleigh distribution. The mean load is given by 
$\bE{L}=L_{\textrm{min}}+ \lambda \Gamma (1+1/k)$, where $\Gamma(\cdot)$ is the gamma-function given by
$\Gamma(x)=\int_{0}^{\infty} t^{x-1} e^{-t} dt$. 
\end{itemize}

First, we confirm our results presented in Sections \ref{subsec:Results_1} and \ref{subsec:Results_2}
concerning the response of the system to attacks of varying size; i.e. concerning the final system size  $n_{\infty}(p)$ under different load-extra space distributions including its transition behavior around the critical attack size $p^{\star}$.  
In all simulations, we fix the number of lines at $N=10^6$, and for each set of parameters being considered (e.g., the distribution $p_{LS}(x,y)$ and attack size $p$) we run 200 independent experiments.  
The results are shown in Figure
\ref{fig:general}
where symbols represent the {\em empirical} value of the final system size $n_{\infty}(p)$ (obtained by averaging over 200 independent runs for each data point), and lines represent the
analytical results computed from (\ref{eq:main_condition}) and (\ref{eq:final_size}). We see that theoretical results match the simulations very well in all cases. 

The specific distributions used in Figure
\ref{fig:general} are as follows: From left to right, we have 
i) $L$ is Weibull with $L_{\textrm{min}}=10,\lambda =100, k=0.4$ and $S=\alpha L$ with $\alpha = 1.74$;
ii) $L$ is Uniform over [10,30] and $S$ is Uniform over [1,5];
iii) $L$ is Weibull with $L_{\textrm{min}}=10,\lambda =10.78, k=6$ and $S$ is Uniform over [5,10];
iv) $L$ is Pareto with $L_{\textrm{min}}=10, b=2$, and $S=\alpha L$ with $\alpha=0.7$;
v) $L$ is Uniform over [10,30] and $S$ is Uniform over [10,60];
and
vi) $L$ is Weibull with $L_{\textrm{min}}=10,\lambda =10.78, k=6$ and $S$ is Uniform over [20,100].
Thus, the plots in Figure
\ref{fig:general} demonstrate the effect of the load-free space distribution on the robustness of the resulting power system. We see that both the {\em family} that the distribution belongs to (e.g., Uniform, Weibull, or Pareto) as well as the specific parameters of the family affect the behavior of $n_{\infty}(p)$. For instance, the curves representing the two cases where $L$ and $S$ follow a Uniform distribution demonstrate that both {\em abrupt} ruptures and ruptures with a preceding divergence are possible in this setting, depending on the parameters $L_{\textrm{min}}, L_{\textrm{max}}, S_{\textrm{min}}$ and $S_{\textrm{max}}$.
In cases where the load follows a Pareto distribution and $S=\alpha L$, only abrupt ruptures are possible as shown in \cite{yaugan2015robustness}.  
Finally, we see that the Weibull distribution gives rise to a richer set of possibilities for the transition of $n_{\infty}(p)$.
Namely, we see that not only we can observe an abrupt rupture, or a rupture with preceding divergence (i.e., a second-order transition followed by a first-order breakdown), it is also possible that $n_{\infty}(p)$ goes through a first-order transition (that does not breakdown the system) followed by a second-order transition that is followed by an ultimate first-order breakdown; see the behavior of the orange circled line in Figure \ref{fig:general}.
We remark that these cases occur when $h(x)$ has a local maximum at $x=S_{min}$, while its global maximum occurs at a later point $x>S_{min}$; see \cite{yaugan2015robustness} for a more detailed discussion of this matter.

\begin{figure}[t] 
  \begin{subfigure}[b]{0.5\linewidth}
    \centering
    \includegraphics[width=0.8\linewidth]{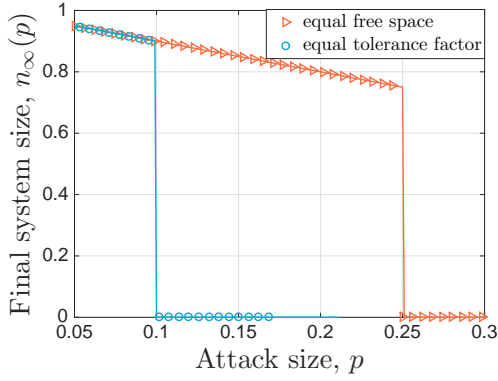} 
    \caption{Load follows Uniform distribution} 
    \label{fig:opt_S_alpha_1} 
    \vspace{4ex}
  \end{subfigure}
  \begin{subfigure}[b]{0.5\linewidth}
    \centering
    \includegraphics[width=0.8\linewidth]{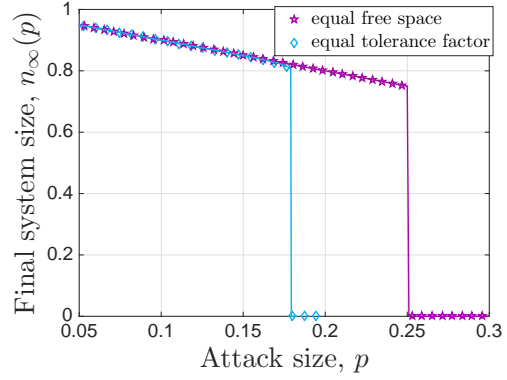} 
    \caption{Load follows Weibull distribution} 
    \label{fig:opt_S_alpha_2} 
    \vspace{4ex}
  \end{subfigure} 
  \begin{subfigure}[b]{0.5\linewidth}
    \centering
    \includegraphics[width=0.8\linewidth]{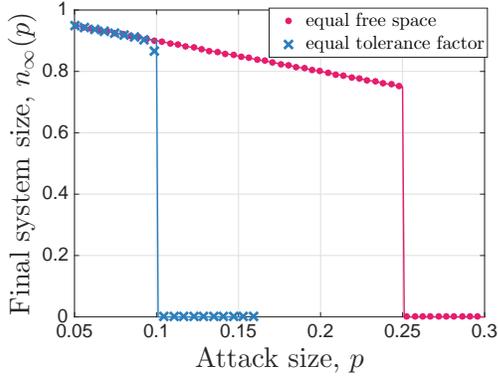} 
    \caption{Load follows Pareto distribution} 
    \label{fig:opt_S_alpha_3} 
  \end{subfigure}
  \begin{subfigure}[b]{0.5\linewidth}
    \centering
    \includegraphics[width=0.8\linewidth]{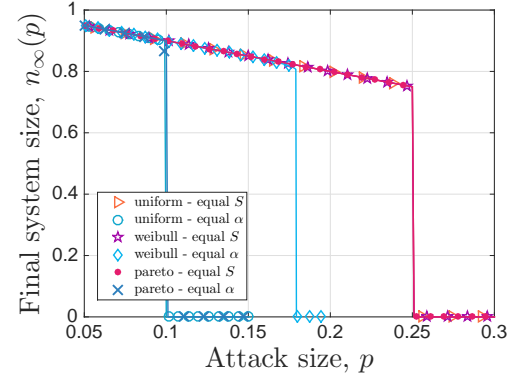} 
    \caption{Comparison of different load distribution} 
    \label{fig:opt_S_alpha_4} 
  \end{subfigure} 
  \caption{ \sl \textbf{Final system size under equal free space vs.~equal tolerance factor.} In all cases, we set $L_{min}=10$,  $\mathbb{E}[L]=30$, and $\mathbb{E}[S]=10$. When load follows Weibull distribution we let $k = 6$ and set $\lambda = 20/\Gamma(1+1/k)$ so that $\mathbb{E}[L]=30$. In each of the three cases, we either let $S \sim \delta(\mathbb{E}[S])$ meaning that all lines have the same free space, or we set $S_i = \alpha L_i$
  with $\alpha = \bE{L}/\bE{S}=1/3$ so that the mean free space still equals 10. We see that analysis (represented by lines) match the simulations (shown in symbols) very well and that robustness is indeed optimized by equal free-space allocation regardless of how initial load is distributed. We also see that system is significantly more robust under equal free space allocation as compared to the case of the equal tolerance factor.}
  \label{fig:opt_S_alpha} 
\end{figure}

In our second set of simulations we seek to verify the results presented in Section \ref{subsec:Results_3}, namely the optimality of assigning the same free space to all lines (regardless of how initial loads are distributed) in terms of maximizing the robustness. In the process, we also seek to compare the robustness achieved under equal free-space distribution versus the commonly used strategy of setting $S_i=\alpha L_i$ for each line. We note that the latter setting with a universal tolerance factor $\alpha$ is commonly used in relevant research literature \cite{MotterLai,WangChen,Mirzasoleiman,CrucittiLatora} as well as in industrial applications \cite{bernstein2014power,kinney2005modeling}; therein, the term $(1+\alpha)$ is sometimes referred to as the {\em Factor of Safety}.
The results are depicted in Figure \ref{fig:opt_S_alpha} where
lines represent the analytical results given in Section \ref{subsec:Results_3} and symbols are obtained by averaging over 200 independent experiments with $N=10^6$ lines. In all cases we
fix the mean load at $\bE{L}=30$ and mean free-space at $\bE{S}=10$. With load distributed as Uniform (Figure \ref{fig:opt_S_alpha_1}), Weibull (Figure \ref{fig:opt_S_alpha_2}), or Pareto (Figure \ref{fig:opt_S_alpha_3}), we either let $S_i=10$ for all lines, or use $S_i = \alpha L_i$ with $\alpha = \bE{S}/\bE{L} = 1/3$, the latter choice making sure that the mean free-space is the same in all plots.

We see in all cases that there is an almost perfect agreement between theory and simulations. 
We also confirm that regardless of how initial load is distributed, the system achieves uniformly optimal robustness (i.e., maximum $n_{\infty}(p)$ for all $p$) 
as long as the free-space is distributed equally; e.g., see Figure \ref{fig:opt_S_alpha_4} that combines all plots in Figures 
\ref{fig:opt_S_alpha_1}-\ref{fig:opt_S_alpha_3}.
In other words, we confirm that (\ref{eq:robustness_of_dirac_S}) holds with 
the critical attack size $p^{\star}$ matching the optimal value $p^{\star}_{optimal}=\bE{S}/\bE{C}=0.25$.
Finally, by comparing the robustness curves under equal free-space and equal tolerance factor, we see the dramatic impact of free-space distribution on the robustness achieved. To give an example, we see from Figure \ref{fig:opt_S_alpha_4} that regardless of how initial load is distributed, the system can be made robust against random attacks that fail up to 25\% of the lines; as already discussed this is achieved by distributing the total free-space equally among all lines. However, if the standard approach of setting the free-space proportional to the initial load is followed, the system robustness can be considerably worse with attacks targeting as low as 10\% of the lines being able to breakdown the system.





\vspace{2mm}
\noindent{\bf Real wold data.} 
\label{subsec:IEEE_data}
Thus far, our analytical results are tested only on synthetic data; i.e., simulations are run when
load-free space variables
$\{L_i, S_i\}_{i=1}^{N}$ are
generated {\em randomly} from commonly known distributions. 
To get a better idea of the real-world implications of our work, we also run simulations on power flow data from the IEEE power system test cases \cite{christieuniversity}; the IEEE test-cases are widely regarded as realistic test-beds and used commonly in the literature. Here, we consider four power flow test cases corresponding to the 30-bus, 57-bus, 118-bus, and 300-bus systems. For each test case, we
take the load values directly from the data-set \cite{christieuniversity}. Since the data-set does not contain the line capacities, we allocate all lines an equal free-space, $S=10$; clearly, most of the discussion here would follow with different free-space distributions. 

\begin{figure}[!t] 
  \begin{subfigure}[b]{0.5\linewidth}
    \centering
    \includegraphics[width=0.8\linewidth]{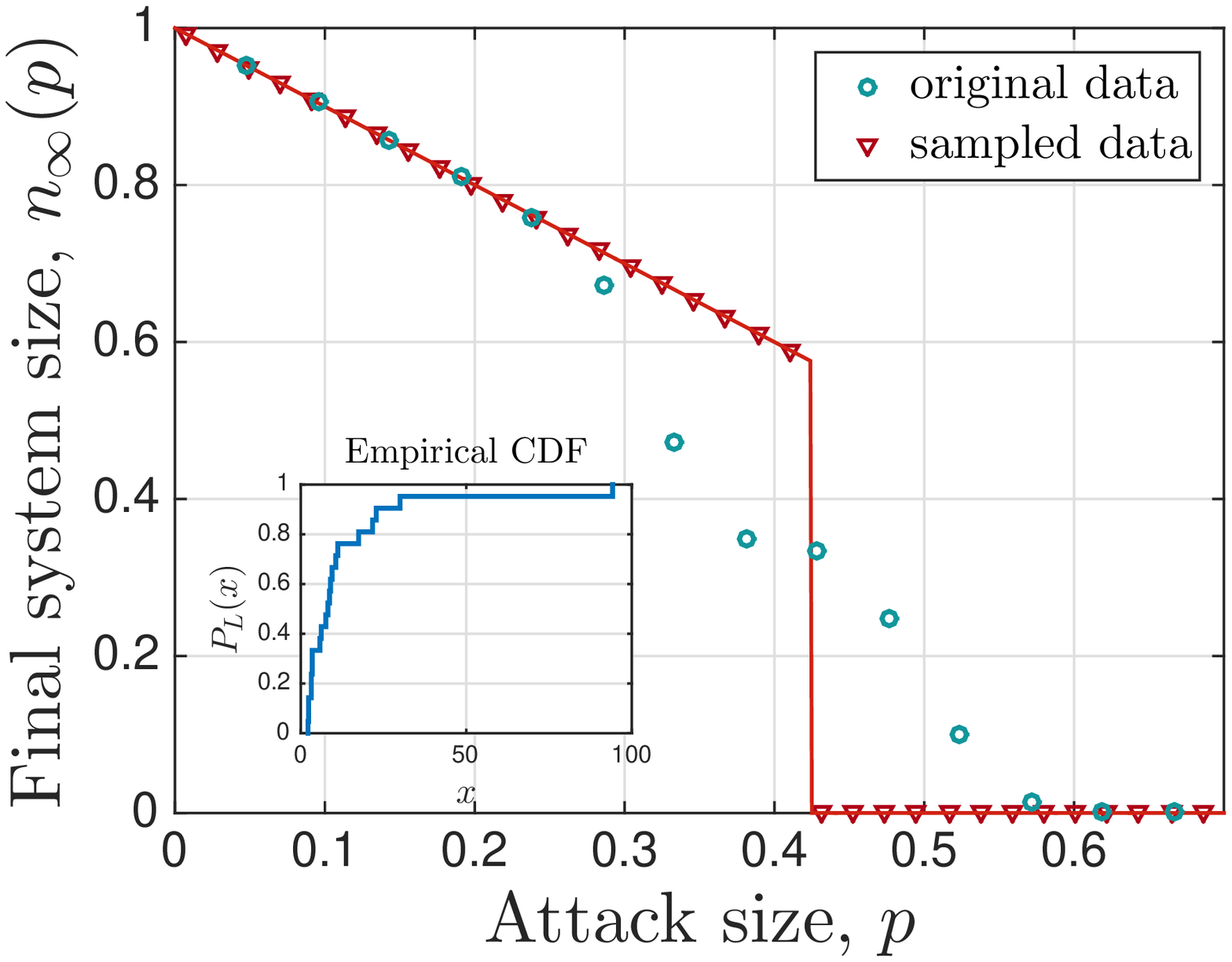} 
    \caption{IEEE 30 bus test case} 
    \label{fig3:a} 
    \vspace{4ex}
  \end{subfigure}
  \begin{subfigure}[b]{0.5\linewidth}
    \centering 
    \includegraphics[width=0.8\linewidth]{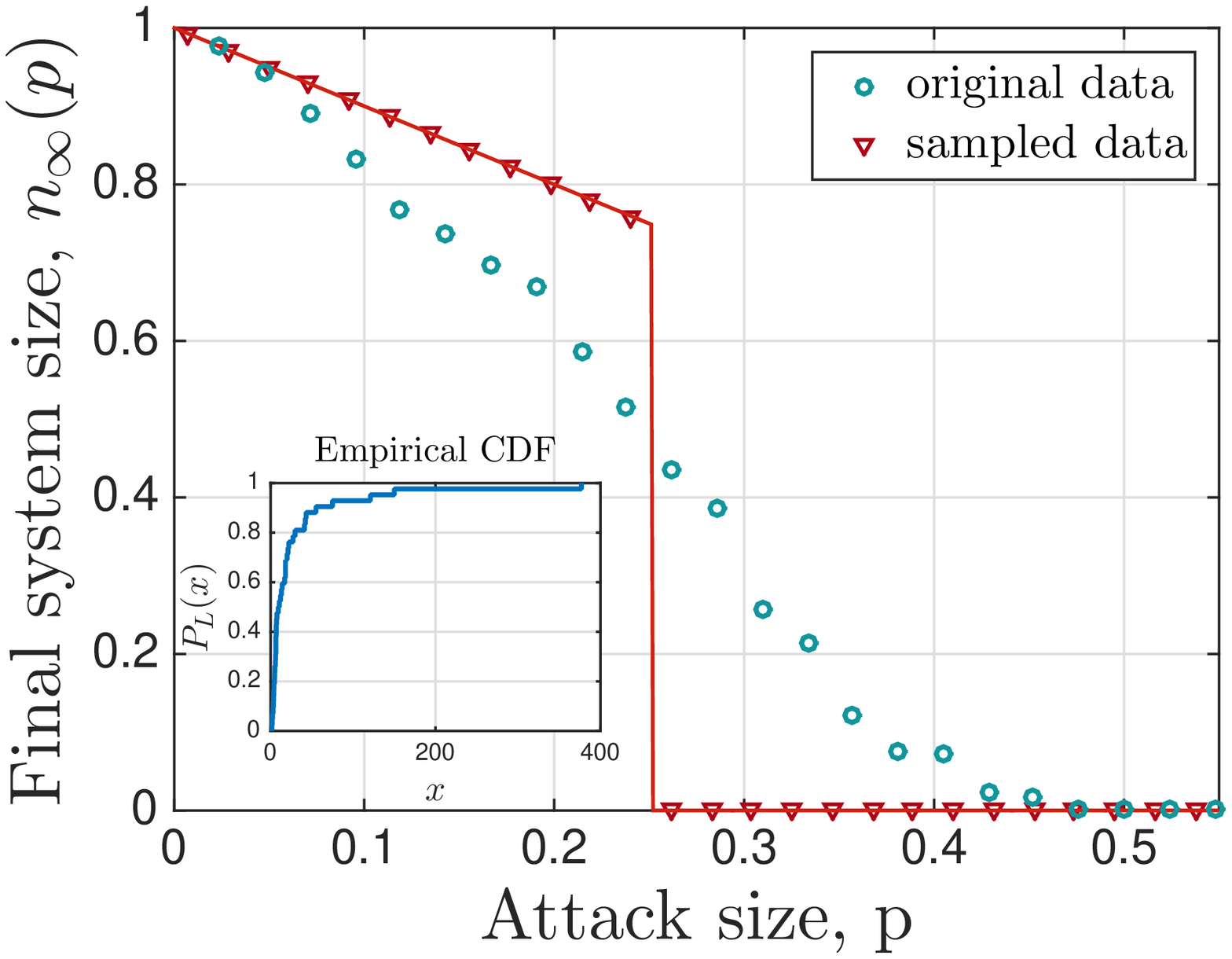} 
    \caption{IEEE 57 bus test case} 
    \label{fig3:b} 
    \vspace{4ex}
  \end{subfigure} 
  \begin{subfigure}[b]{0.5\linewidth}
    \centering
    \includegraphics[width=0.8\linewidth]{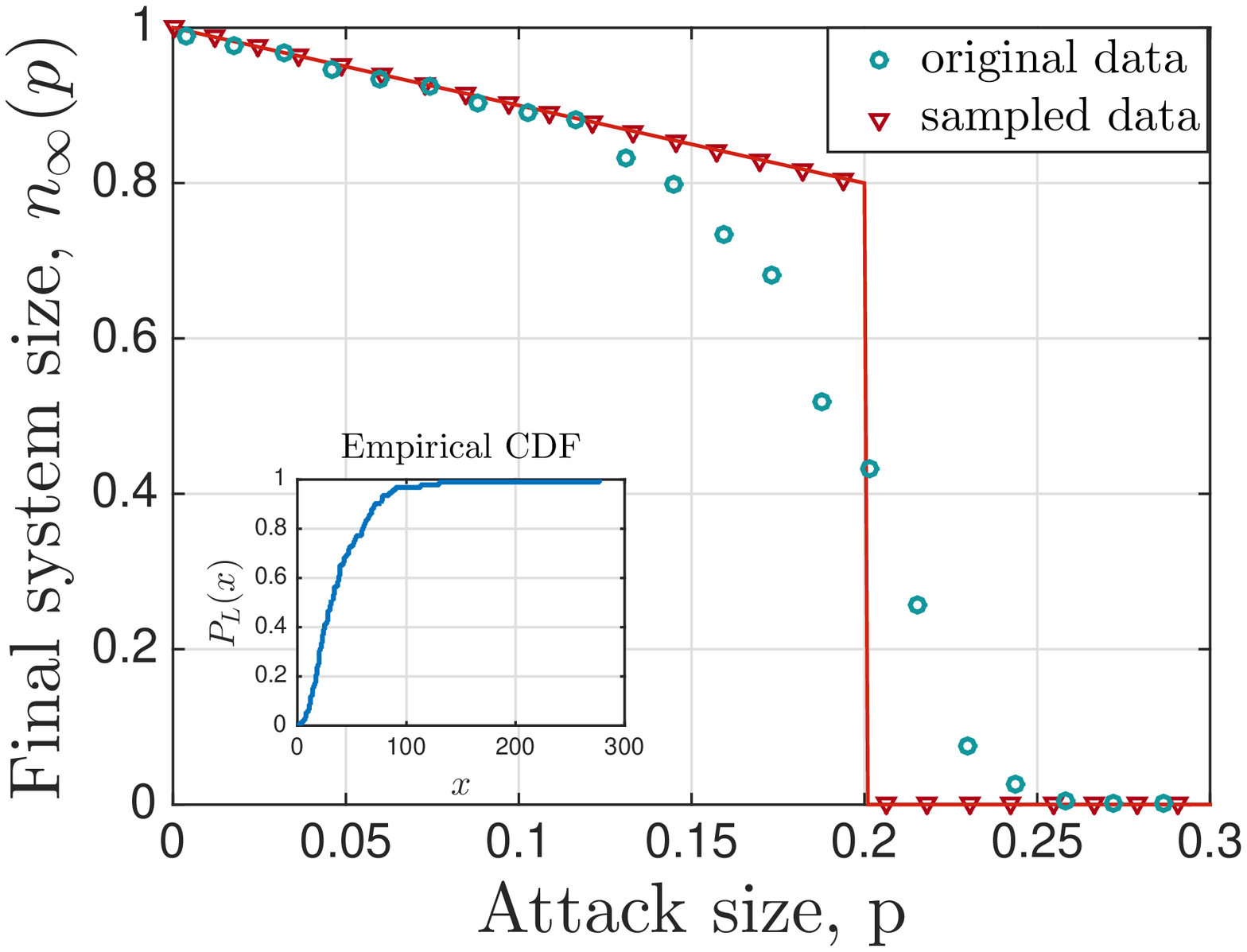} 
    \caption{IEEE 118 bus test case} 
    \label{fig3:c} 
  \end{subfigure}
  \begin{subfigure}[b]{0.5\linewidth}
    \centering
    \includegraphics[width=0.8\linewidth]{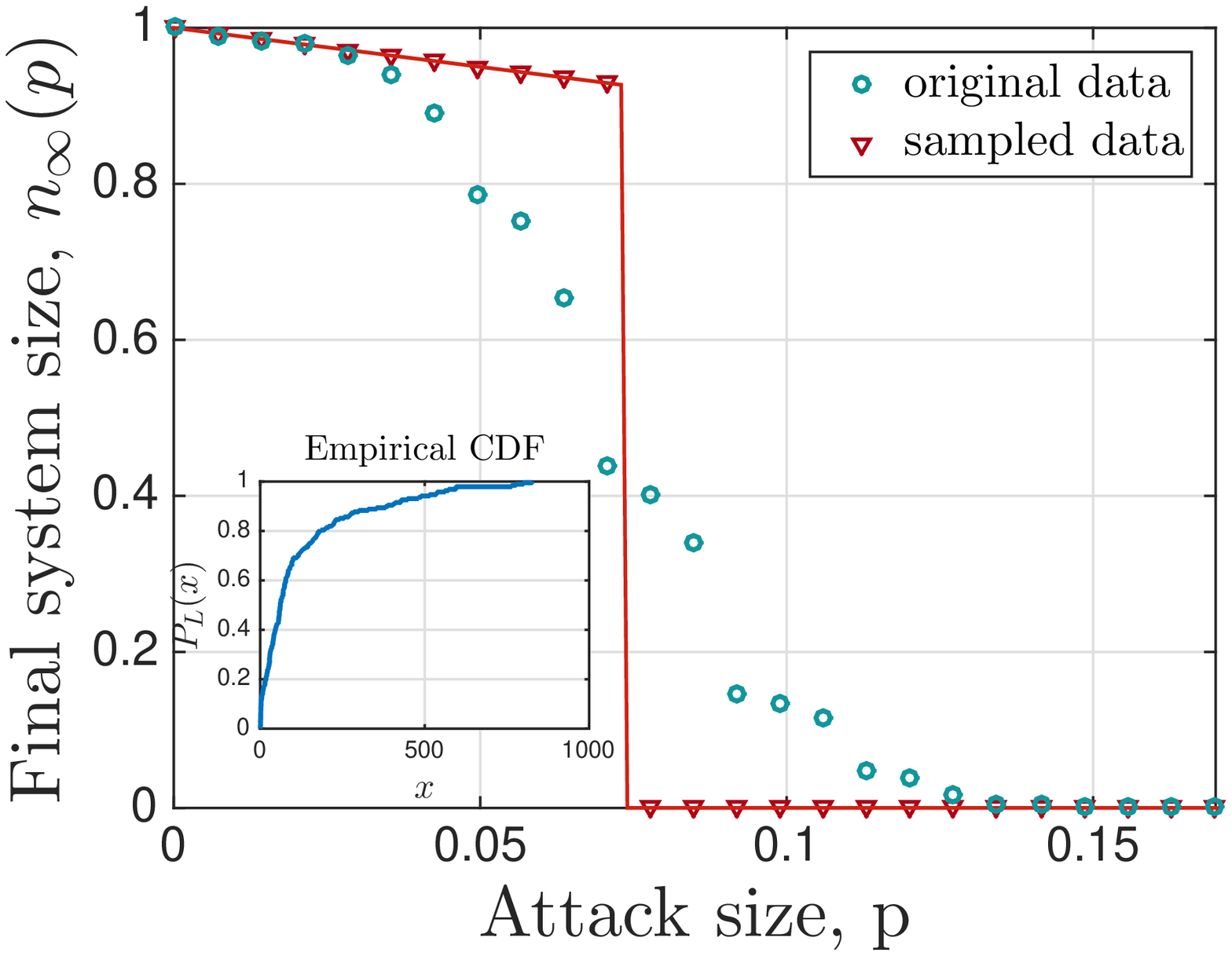} 
    \caption{IEEE 300 bus test case} 
    \label{fig3:d} 
  \end{subfigure} 
  \caption{\sl \textbf{Simulation results on IEEE test cases.} The initial load values are  taken directly from the corresponding IEEE test-case data-sheet \cite{christieuniversity}, and each line is given an equal free space of
$\mathbb{E}[S]=10$. The empirical distribution of load is shown in the Inset of each figure, and the mean load values are given by 
 13.54,  29.95, 39.95, and 125.02 for the  30-bus system, 57-bus system,
 118-bus system, and 300-bus system, respectively. The blue circles represent the simulation results for the final system size $n_{\infty}(p)$. The theoretical results (shown in lines) capture the overall tendency of $n_{\infty}(p)$ but fail to predict the numerical results well, especially around the critical attack size. We see that this is merely a finite-size effect as we sample $N=10^5$ load values from the empirical distribution and repeat the same experiment. The results are shown in red triangles and are in perfect agreement  with the analysis.}
  \label{fig:ieee_data} 
\end{figure}

Figure \ref{fig:ieee_data} presents the results from the IEEE data set simulations, where blue circles represent the final system size $n_{\infty}(p)$ under original load data from each test case; each data point is obtained by averaging the result of  200 independent random attack experiments. 
As we compare these circles with our analytical results (represented by solid red lines) we see that the overall tendency of $n_{\infty}(p)$ is in
accords with the theoretical analysis. However, the agreement of theory and simulations is significantly worse than that observed in Figures \ref{fig:general} and \ref{fig:opt_S_alpha}. 
This is because our mean field analysis relies on the number of lines $N$ being {\em large}, while the IEEE test case data represent very small systems; e.g., the underlying systems have 30, 57, 118, and 300 lines in Figures \ref{fig3:a}-\ref{fig3:d}, respectively. 
In order to verify that the mismatch is due to the small system size (rather than the load distribution being different from commonly known ones), we re-sample $10^5$ load values from the {\em empirical} load distribution obtained from the data-set in each case; the Inset in each figure shows the corresponding empirical distribution $P_L(x)$. The simulation results with these $N=10^5$ load values are shown in  Figure \ref{fig:ieee_data} with red triangles. This time with the number of lines increased, we obtain a perfect match between analysis and simulations. This confirms our analysis under 
realistic load distributions as well. 
We also see that although analytical results fail to match the system robustness perfectly when $N$ is very small, they still capture the overall tendency of the robustness curves pretty well. In fact, they can be useful in predicting attack sizes that will lead to a {\em significant} damage to the system; e.g., in all cases we see that the analytically predicted critical attack size $p^{\star}$, ranging from 0.42 in Figure \ref{fig3:a} to 0.07 in Figure \ref{fig3:d}, leads to the failure of more than 50 \% of all lines in the real system.

\section{Discussion}

Our results provide a complete picture of the robustness of power systems against random attacks under the equal load-redistribution model. Namely, with initial load $L_i$ and extra space $S_i$ of each line being independently and identically distributed with $p_{LS}(x,y)$, our analysis explains how the final system size $n_{\infty}(p)$ will behave under attacks with varying size $p$. We also demonstrate different types behavior that $n_{\infty}(p)$ can exhibit near and around the {\em critical} attack size $p^{\star}$, i.e., the point after which $n_{\infty}(p)=0$ and the system breaks down completely. We show that the final breakdown of the system is always first-order (i.e., discontinuous) but depending on $p_{LS}(x,y)$, this may i) take place abruptly meaning that $n_{\infty}(p)$ follows the $1-p$ line until its sudden jump to zero; or ii) 
be preceded by a second-order (i.e., continuous) divergence from the $1-p$ line. We also demonstrate the possibility of richer behavior where $n_{\infty}(p)$ drops to zero through a first-order, second-order, and then a first-order transition. 
The discontinuity of the final system size at $p^{\star}$ makes it very difficult to predict system behavior (in response to attacks)
from previous data. In fact, this is reminiscent of the real-world phenomena of unexpected large-scale system collapses; i.e., cases where seemingly identical attacks/failures lead to entirely different consequences. 
On the other hand, the cases that exhibit a preceding second-order transition are less severe, since the deviation from the $1-p$ line
may be taken as an early warning that the current attack size is close to $p^{\star}$ and that the system is not likely to sustain attacks much larger than this.

From a design perspective, it is desirable to maximize the robustness of the electrical power system under certain constraints. In our analysis, we address this problem and derive the 
optimal load-free space distribution $p_{LS}(x,y)$ that maximizes the final system size $n_{\infty}(p)$ uniformly for all attack sizes $p$. Namely, we show that under the constraints that $\bE{L}$ and $\bE{S}$ are fixed, robustness is maximized by allocating the the same free space to all lines and distributing the initial loads arbitrarily; i.e. the distribution $p_{LS}(x,y) = p_L(x)\delta(y-\mathbb{E}[S])$ maximizes robustness for arbitrary $p_L(x)$. We show that this optimal distribution leads to significantly better robustness than the commonly used strategy of assigning a universal tolerance factor $\alpha$, i.e., using $p_{LS}(x,y) = p_L(x)\delta(y-\alpha x)$.


Our theoretical results are verified via extensive simulations using both synthetic data and real world data. We show that our results are in perfect agreement with numerical simulations when the system size $N$ is large; in most cases it suffices to have $N=10^4$ to $N=10^5$. However, we see from our simulations with the IEEE test-cases that when $N$ is very small (we considered $N=30$, $N=57$, $N=118$, and $N=300$), our theory fails to yield the same prediction accuracy. Nevertheless, we see that our results  
capture the overall tendency of $n_{\infty}(p)$ pretty well, and thus can serve as a useful predictor of the critical attack size.

An important direction for future work would be to relax the simplifications and assumptions used here for modeling the failures in an electrical power system. For example, the equal redistribution rule is used here to capture the long-range effect of the Kirchhoff Law, i.e., that the failure of a line may impact the system {\em globally}. Future work may consider different types of {\em global} load redistribution rules that are not based on {\em equal} redistribution; e.g., load may be redistributed randomly or according to some other  rules. Hybrid approaches where a fraction of the load is redistributed only locally to the neighboring lines, while the rest being redistributed globally might be considered. 
Another interesting direction for future work would be to consider the case of {\em targeted} attacks, rather than random attacks studied here. A good starting point in that direction would be to study possible attack strategies that a capable adversary might use; e.g., given $L_1,\ldots, L_N$ and $S_1,\ldots,S_N$, which $k$ lines should an adversary attack in order to minimize the final system size $n_{\infty}$? A preliminary analysis of this problem can be found in \cite{Evangelos}, with partial results indicating that optimal attack strategies may be computationally expensive to derive -- i.e., that the problem is NP-Hard. 

\section{Methods}

\subsection{Understanding the cascade dynamics}
\label{subsec:methods_1}
Our proofs are based on a mean-field analysis of the cascading failure dynamics under the equal redistribution model; see Model Definitions for details.
Assume that failures take place in discrete time steps $t=0,1,\ldots$, and are initiated at time $t=0$ by the random failure of a $p$-fraction of the lines.
For each $t=0,1,\ldots$, let $f_t$ denote the fraction of lines that have {\em failed} until stage $t$. The number of links that are still alive at time $t$ is then given by $N_t = N (1-f_t)$; e.g., $f_0=0$ and $N_0 = N (1-p)$. Also, we find it useful to denote by $Q_t$ the total {\em extra} load per {\em alive} line at (the end of) stage $t$. In other words, $Q_t$ is given by the total load of all $f_t N$ failed lines until this stage divided by $(1-f_t)N$; e.g., $Q_0 = p \bE{L}/(1-p)$ since the initial attack is {\em random}.

Our main goal is to derive the final system size $n_{\infty}(p)$ as a function of the attack size $p$. With the above definitions in place, we clearly have 
$n_{\infty}(p)=1-f_{\infty}$.  Thus, the derivation of $n_{\infty}(p)$ passes through an understanding of the behavior of $f_t$ as $t \to \infty$. Here, we will achieve this by first  deriving recursive relations for $f_t, Q_t$, and $N_t$ for each $t=0,1, \ldots$, and then analyzing the steady-state behavior of the recursions.
This method has already proven successful by Ya\u{g}an \cite{yaugan2015robustness}, who studied the same problem  in a special case where
\[
S_i = \alpha L_i, \quad i=1,\ldots, N
\]
with $\alpha>0$ defining the {\em universal} tolerance factor. Put differently, the work \cite{yaugan2015robustness} considers the special case where
\[
p_{LS}(x,y) = p_L(x) \delta(y-x\alpha)
\]
for arbitrary distribution of initial load $p_L(x)$. Here, we start our discussion from the recursions derived by Ya\u{g}an \cite[Eqn. 6]{yaugan2015robustness}
for this special case. Namely, with $f_0=p$, $Q_0=p \bE{L}/(1-p)$ (and $Q_{-1}=0$), they showed for each $t=1,2,\ldots$
 that 
\begin{align} \label{eq:1}
f_{t+1} &= 1 - (1-f_t) \mathbb{P}[\alpha L > Q_t \Bigm| \alpha L>Q_{t-1}]
\\  \label{eq:2}
Q_{t+1} &= \frac{p\mathbb{E}[L]+(1-p)\mathbb{E}[L \cdot \mathbbm{1}[\alpha L \leq Q_t]]}{(1-p)\bP{\alpha L > Q_t}}
\\ \label{eq:3}
N_{t+1} &= (1-f_{t+1})N
\end{align} 

An inspection of their derivation reveals that these recursive relations do hold in the general case as well with $\alpha L$ replaced by the  random variable $S$. 
Namely, with no constraints imposed on the distribution $p_{LS}(x,y)$ (other than those stated in Model Definitions), we have
\begin{align} \label{eq:1b}
f_{t+1} &= 1 - (1-f_t) \mathbb{P}[S > Q_t \Bigm| S > Q_{t-1}]
\\  \label{eq:2b}
Q_{t+1} &= \frac{p\mathbb{E}[L]+(1-p)\mathbb{E}[L \cdot \mathbbm{1}[S \leq Q_t]]}{(1-p)\bP{S>Q_t}}
\\ \label{eq:3b}
N_{t+1} &= (1-f_{t+1})N
\end{align} 
These equations can also be validated intuitively. First of all, given that $Q_t$ defines the extra load per alive line at the end of stage $t$, we know that for a line 
to fail exactly at stage $t+1$, it must have a free space smaller than $Q_t$ but larger than $Q_{t-1}$, with the latter condition ensuring that the line does not fail at any previous stages. So, the fraction of lines that fail at stage $t+1$ among those that survive stage $t$ is intuitively given by $\bP{S \leq Q_t ~| S > Q_t}$. 
Rewriting  (\ref{eq:1b}), we get
\[
\frac{\textrm{$\#$  of lines that survive stage $t$ but fail at $t+1$}}{\textrm{$\#$ of lines that survive stage $t$} } = \frac{f_{t+1} - f_t}{1-f_t} =
1-\frac{1-f_{t+1}}{1-f_t} =1 - \mathbb{P}[S > Q_t \Bigm| S > Q_{t-1}] = \mathbb{P}[S \leq Q_t \Bigm| S > Q_{t-1}] 
\]
confirming this intuitive argument. In fact, it is clear that for a line to survive stage $t+1$, it must i) survive the initial attack (which happens with probability $1-p$), and ii) have a free space $S > Q_t$. Given the independence of the initial attack from other variables, we thus have $1-f_{t+1} = (1-p) \bP{S>Q_t}$.  
This explains the denominator of (\ref{eq:2b}) since $Q_{t+1}$ gives the additional load {\em per alive} line at this stage. The nominator of (\ref{eq:2b}) should then give the mean total load of the lines that have failed until (and including) stage $t+1$, normalized with $N$; the normalization is required given that the denominator term is also normalized with $N$. To calculate the total load of the failed lines at this stage, first we note that $p$ fraction of the lines fail randomly as a result of the initial attack, giving the term $p\bE{L}$ in the nominator of  (\ref{eq:2b}). In addition, among the remaining $1-p$ fraction of the lines, those with free space satisfying $S \leq Q_t$ will fail, leading to the second term in the nominator of  (\ref{eq:2b}).

Returning to the recursions (\ref{eq:1})-(\ref{eq:3}), we see that cascading failures will stop and a steady-state will be reached when $f_{t+2}=f_{t+1}$. From (\ref{eq:1b}), we see that this occurs if 
\[
\bP{S > Q_{t+1} ~|~ S > Q_t} =1,
\]
or, equivalently if
\begin{align}
 &  \bP{S >  \frac{p\mathbb{E}[L]+(1-p)\mathbb{E}[L \cdot \mathbbm{1}[S \leq Q_t]]}{(1-p)\bP{S>Q_t}} ~\Bigg|~ S > Q_{t}} 
= 1,
 \label{eq:stop_condition} 
\end{align}
upon using (\ref{eq:2b}). In order to simplify this further, we let $x:=Q_t$, and realize that
\begin{align}\nonumber
 p \bE{L}  + (1-p) \bE{L \cdot \1{S \leq  x}} 
 & =p \bE{L}  + (1-p) \bE{L \cdot (1-\1{S >  x})} 
  = \bE{L} - (1-p) \bE{L \cdot \1{S >  x}}.
\end{align}
With these in place, the condition for cascades to stop (\ref{eq:stop_condition}) gives 
\begin{equation} \label{eq:19}
\bP{ S>\frac{ \bE{L}-(1-p) \bE{L \cdot \mathbbm{1}[S > x]}}{(1-p)\mathbb{P}[S > x]} ~ \Bigg| ~ S > x }=1 \qquad (\textrm{\bf Cascade stop condition})
\end{equation}

It is now clear how to obtain the fraction $n_{\infty}(p)$ of power lines that are still alive at the end of the cascading failures. 
First, we shall find the smallest solution $x^{\star}$ of (\ref{eq:19}) that gives the equilibrium value $Q_{\infty}$ at which cascades will stop.  
Then the final-fraction $n_{\infty}$ of alive lines is given by
\begin{equation} \label{eq:20}
n_\infty (p) = 1-f_\infty = (1-p)\mathbb{P}[ S > x^{\star}] 
\end{equation} 
The last relation follows from the fact that $1-f_{t+1} = (1-p) \bP{S > Q_t}$ for each $t=0,1,\ldots$. This can be established 
in the following manner.
Applying (\ref{eq:1b}) repeatedly, we get
\begin{eqnarray}\nonumber
\begin{array}{ll}
1-f_{t+1} &= (1-f_t) \bP{S > {Q_t} ~|~ S > {Q_{t-1}}} \\
1-f_{t} &= (1-f_{t-1}) \bP{S > {Q_{t-1}} ~|~ S > {Q_{t-2}}} \\
~~~\vdots & \\
1-f_{1} &= (1-f_0) \bP{S > Q_0}, 
\end{array}
\end{eqnarray}
which gives
\[
1-f_{t+1}  = (1-f_0) \prod_{\ell = 0}^{t} \bP{S > {Q_{\ell}} ~\big|~ S > Q_{\ell-1}},
\]
where $Q_{-1} = 0$ as before. Since $Q_t$ is monotone increasing in $t$, i.e., $Q_{t+1}\geq Q_t$ for all $t$, we get
\begin{align}
1-f_{t+1}  =  (1-f_0) \frac{\bP{S > Q_{t}}}{\bP{S > Q_{t-1}}} \cdot \frac{\bP{S > Q_{t-1}}}{\bP{S > Q_{t-2}}} \cdots \frac{\bP{S > Q_{1}}}{\bP{S > Q_{0}}} 
  \cdot \bP{S > Q_{0}}
  = (1-p)\bP{S > Q_{t}} 
  \label{eq:simplified_f_t}
\end{align}
as we recall that $f_0=p$. 

We now seek to simplify the cascade stop condition given at (\ref{eq:19}). 
For notational convenience, let 
\begin{align}
g(x) := \frac{E[L]-(1-p)E[L \cdot \mathbbm{1}[S > x]]}{(1-p)\mathbb{P}[S > x]}.
\label{eq:defn_g_x}
\end{align}
Then, (\ref{eq:19}) becomes
\begin{equation} \label{eq:21}
\bP{S > g(x) ~\big|~ S > x}=1,
\end{equation}
which holds if either one of the following is satisied:
\begin{enumerate}
\item $x \geq g(x)$; or,
\item $x < g(x)$ and $\bP{S > g(x) ~\big|~ S > x}=1$.
\end{enumerate}
The next result (proved in the Supplementary Material) shows that it suffices to consider only the first case for the purposes of our discussion.
\begin{claim}
Let $x^{\star}$ be the smallest solution of (\ref{eq:21}), and $x^{\star \star}$ be the smallest solution of 
$x \geq g(x)$. Then, we have
\[
\bP{S > x^{\star}} = \bP{S > x^{\star \star}}
\]
\label{claim:1}
\vspace{-5mm}
\end{claim}
The proof of this important technical result is given in the Supplementary File.

Rewriting the inequality $x \geq g(x)$ and using Claim \ref{claim:1}, we now establish the first main result of the paper given at (\ref{eq:main_condition}) and 
(\ref{eq:final_size}). Namely, with $x^{\star}$ denoting the smallest solution of
\begin{equation} \label{eq:23}
\mathbb{P}[S>x]( x+\mathbb{E}[L \mid S>x]) \geq \frac{\mathbb{E}[L]}{1-p}, \qquad x \in (0, \infty)
\end{equation}
the final system size $n_{\infty}(p)$ is given by $(1-p)\bP{S>x^{\star}}$. If (\ref{eq:23}) has no solution, we set $x^{\star}=\infty$ leading to $n_{\infty}(p)=0$.
\myendpf

\noindent{\bf Critical attack size.}
Characterizing the critical attack size $p^{\star}$ is now a simple matter from the discussion above. 
Recall that critical attack size is defined via
\[
p^{\star} = \sup\{p: n_{\infty}(p) > 0 \}.
\]
Since $\bE{L}/(1-p)>0$ always holds, we know that whenever there exists an $x^{\star}<\infty$ satifying (\ref{eq:23}), we must have that 
$\mathbb{P}[S>x^{\star}] > 0$ as otherwise the function $g(x^{\star})=\mathbb{P}[S>x^{\star}]( x^{\star}+\mathbb{E}[L \mid S>x^{\star}])$ would be zero conflicting with (\ref{eq:23}).
Therefore, we have $n_{\infty}(p)>0$ for any $p$ for which (\ref{eq:23}) has a solution; and we get $n_{\infty}(p) = 0$
only if (\ref{eq:23}) has no solution. With these in mind, it is clear that $p^{\star}$ will be given by the supremum of $p$ values for which (\ref{eq:23}) has a solution. Equivalently, $p^{\star}$ is given by the value of $p$ for which $\bE{L}/(1-p)$ equals to the {\em global} maximum of $g(x)$. More precisely, we have
\begin{equation} \label{eq:26}
p^{\star}=1-\frac{\mathbb{E}[L]}{\max \limits_{x} \{\mathbb{P}[S>x]( x+\mathbb{E}[L \mid S>x])\}}
\end{equation}
This establishes (\ref{eq:max_attack}).
\myendpf

\subsection{Order of phase transition and condition for abrupt break-downs}
\vspace{1mm}
\noindent{\bf Order of transition.}
We now establish the fact that the final breakdown of the system will always be through a {\em first-order} (i.e., discontinuous) transition. 
This amounts to establishing (\ref{eq:first_order}), namely that $n_{\infty}(p^{\star})>0$; this then implies a discontinuity in $n_{\infty}(p)$ at $p=p^{\star}$, since by definition of the critical attack size we have $n_{\infty}(p^{\star} + \epsilon) = 0$ for any $\epsilon>0$. 

We now establish $n_{\infty}(p^{\star})>0$. From (\ref{eq:26}), we see that when $p=p^{\star}$ the cascade stopping condition (\ref{eq:23}) will be satisfied
by $x_{\star}$ that maximizes $\mathbb{P}[S>x]( x+\mathbb{E}[L \mid S>x])$. In other words, we have
\begin{equation}
n_{\infty}(p^{\star}) = (1-p^{\star})\bP{S>x_{\star}} 
\label{eq:final_size_at_criticality}
\end{equation}
where
\[
x_{\star} = \arg \max \limits_{x} \{\mathbb{P}[S>x]( x+\mathbb{E}[L \mid S>x])\}
\]
We argue that $\bP{S>x_{\star}} > 0$ by contradiction. Suppose that $\bP{S>x_{\star}} = 0$. With $x_{\star}$ denoting the global maximizer of $\mathbb{P}[S>x]( x+\mathbb{E}[L \mid S>x])$, this would imply that 
\begin{equation}
\mathbb{P}[S>x]( x+\mathbb{E}[L \mid S>x]) \leq \mathbb{P}[S>x_{\star}]( x_{\star}+\mathbb{E}[L \mid S>x_{\star}]) = 0, \qquad \textrm{for all}~~ x \in (0, \infty).
\label{eq:contradiction}
\end{equation}
However, 
with $0 \leq x < S_{\textrm{min}}$, we have $\mathbb{P}[S>x]( x+\mathbb{E}[L \mid S>x]) = x + \bE{L}>0$ contradicting with (\ref{eq:contradiction}). Therefore, we must have $\bP{S>x_{\star}} > 0$, and the desired conclusion $n_{\infty}(p^{\star})>0$ immediately follows. \myendpf

\noindent{\bf Condition for an {\em abrupt} first-order transition.} An abrupt rupture is said to take place if the linear decay of $n_{\infty}(p)$ (in the form of $1-p$) is followed by a sudden discontinuous jump to zero at $p^{\star}$; i.e., when it holds that 
\begin{equation}
n_{\infty}(p) = \left\{ 
\begin{array}{cc}
1- p  & \textrm{if $p \leq p^{\star}$}     \\
 0   &      \textrm{if $p > p^{\star}$}
\end{array}
\right.
\label{eq:abrupt_rupture_defn}
\end{equation}
We now show that this occurs if and only if the function $h(x)=\mathbb{P}[S>x]( x+\mathbb{E}[L \mid S>x])$ takes its global maximum at $x_{\star}=S_{\textrm{min}}$. First of all, $n_{\infty} (p^{\star}) = 1 - p^{\star}$ implies that $\bP{S > x_{\star}} = 1$ in view of (\ref{eq:final_size_at_criticality}). This immediately gives 
$x_{\star} \leq S_{\textrm{min}}$. For the reverse direction, observe that
$h(x)$ is linearly increasing on the range $0 \leq x \leq S_{\textrm{min}}$. In fact, we have 
\begin{equation}
h(x) = x + \bE{L}, \qquad 0 \leq x \leq S_{\textrm{min}}.
\label{eq:h_x_till_S_min}
\end{equation}
Therefore, $h(x)$ reaches at least a local maximum at $x=S_{\textrm{min}}$, implying that $x_{\star} \geq S_{\textrm{min}}$. 
Collecting, we establish $x_{\star} = S_{\textrm{min}}$ as a {\em necessary} condition for an abrupt rupture (i.e., (\ref{eq:abrupt_rupture_defn}))
to take place. Next we show that this is also a sufficient condition. If $x_{\star} = S_{\textrm{min}}$ meaning that $h(x)$ is maximized at $S_{\textrm{min}}$,
then the minimum solution for the inequality $h(x) \geq \bE{L}/(1-p)$, when exists, will appear at $x^{\star}$ that satisfies $x^{\star} \leq S_{\textrm{min}}$. This implies that the final system size $n_{\infty}(p) = (1-p) \bP{S>x^{\star}}$ is either $1-p$ (when a solution to the inequality exists), or zero (when no solution to 
 $h(x) \geq \bE{L}/(1-p)$ exists). In other words, when $x_{\star}=S_{\textrm{min}}$ the final system size will be of the form (\ref{eq:abrupt_rupture_defn}), i.e., an abrupt rupture will take place. 
Combining, we conclude that (\ref{eq:abrupt_rupture_defn}) takes place 
if and only if
\begin{equation}
x_{\star} = \arg \max_{x > 0} \left\{\bP{S > x} \left(x + \bE{L ~|~ S > x} \right) \right\} = S_{\textrm{min}}. \qquad \textrm{\bf (necessary {\em and} sufficient condition for abrupt rupture)}
\label{eq:abrupt_condition_repeat}
\end{equation}

In the Supplementary File, we explore this issue further and provide a {\em necessary} (but not sufficient) condition 
for (\ref{eq:abrupt_condition_repeat}) to hold.
This leverages the fact that for the maximum of $h(x)$ to take place at $x=S_{\textrm{min}}$, the derivative of $h(x)$ must change its sign to negative at this point; this would ensure a {\em local} maximum of $g(x)$ take place at $S_{\textrm{min}}$.  
The resulting necessary condition, given here for convenience, is
\[
 p_S(S_{\textrm{min}})(S_{\textrm{min}}  + \bE{L ~|~ S=S_{\textrm{min}}}) > 1.  \qquad \textrm{\bf (necessary condition for abrupt rupture)}
\]

\subsection{Optimal $P_{LS}$ distribution that maximize robustness}
We now seek to find the optimal load-free space distribution $P_{LS}(x,y)$ that maximizes the robustness of the power system, when
$\bE{L}$ and $\bE{S}$ are fixed. First, we focus on maximizing the critical attack size $p^{\star}$.
 Recall (\ref{eq:26})
and observe that
\begin{align}
h(x) &=\mathbb{P}[S>x]( x+\mathbb{E}[L \mid S>x]) 
\nonumber \\ 
& = x\mathbb{P}[S>x]+\mathbb{E}[L \cdot \mathbbm{1}[ S>x]] 
\nonumber \\
\label{eq:28} & \leq \mathbb{E}[S] + \mathbb{E}[L \cdot \mathbbm{1}[ S>x]] 
\\
& \leq  \mathbb{E}[S] + \mathbb{E}[L]
\nonumber
\end{align}
where we use the Markov inequality at (\ref{eq:28}), i.e. that $\mathbb{P}[S>x] \leq \mathbb{E}[S]/x$. Reporting this into (\ref{eq:26}) we get
\begin{align} \label{eq:29}
p^{\star} &=  1-\frac{\mathbb{E}[L]}{\max \limits_{x} \{h(x)\}} 
 \leq 1- \frac{\mathbb{E}[L]}{\mathbb{E}[S] + \mathbb{E}[L]} =\frac{\mathbb{E}[S]}{\mathbb{E}[S] + \mathbb{E}[L]} = \frac{\bE{S}}{\bE{C}}.
\end{align}
This means that regardless of our choice of the joint distribution $P_{LS}(x,y)$ the critical attack size can never be larger than 
$\frac{\bE{S}}{\bE{C}}$.
 
Next, we show that this upper bound is in fact achievable by a {\em Dirac}-delta distribution of free space $S$. 
Assume that 
\begin{equation}
P_{LS}(x,y) = P_L(x) \1{y \leq \bE{S}}
\label{eq:dirac_osy}
\end{equation}
where the load distribution $P_L(x)$ is  arbitrary; this is equivalent to having
$p_{LS}(x,y)= p_{L}(x) \delta(y - \bE{S})$. 
Let $p^{\star}_{dirac}$ denote the corresponding critical attack size. 
With $x=\bE{S}^-$, we have $\bP{S>x}=1$, and $\mathbb{E}[L \cdot \mathbbm{1}[S>x]]=\mathbb{E}[L]$, so that
\[
\lim_{x\uparrow \mathbb{E}[S]} h(x) =\lim_{x\uparrow \mathbb{E}[S]}   \left(\mathbb{P}[S>x]( x+\mathbb{E}[L \mid S>x]) \right) = \mathbb{E}[S] + \mathbb{E}[L]
\] 
Thus, we have
$\max \limits_{x} \{h(x)\} \geq  \mathbb{E}[S] + \mathbb{E}[L]$, which immediately gives
\begin{align} 
p^{\star}_{dirac} &=1-\frac{\mathbb{E}[L]}{\max \limits_{x} \{h(x)\}}  \geq 1-\frac{\mathbb{E}[L]}{ \mathbb{E}[S] + \mathbb{E}[L]} =\frac{\mathbb{E}[S]}{\mathbb{E}[S] + \mathbb{E}[L]}.
\end{align}
Since the lower bound (\ref{eq:29}) holds for any distribution, we conclude that
\begin{equation} \label{eq:30}
p^{\star}_{dirac} =\frac{\mathbb{E}[S]}{\mathbb{E}[S] + \mathbb{E}[L]}:=p^{\star}_{optimal}.
\end{equation}
This shows that a degenerate distribution on the extra space $S$ leads to optimal (i.e., maximum) critical attack size $p^{\star}_{optimal}$. 
\myendpf

We now show that a Dirac-delta distribution of free space not only maximizes $p^{\star}$ but it maximizes the final system size $n_{\infty}(p)$ uniformly across all attack sizes. It is clear that $n_{\infty}(p) \leq 1-p$ for any distribution $P_{LS}(x,y)$ and any attack size $p$. Thus, our claim will be established if we show that the Dirac-delta distribution (\ref{eq:dirac_osy}) leads to an {\em abrupt} rupture and thus the resulting final system size is in the form given at (\ref{eq:abrupt_rupture_defn}). More precisely, we will get the desired result
\begin{equation}
n_{\infty,dirac}(p) = \left\{ 
\begin{array}{cc}
1- p,  & \textrm{if $p \leq p^{\star}_{optimal}$}     \\
 0,   &      \textrm{if $p > p^{\star}_{optimal}$}
\end{array}
\right.
\label{eq:optimal_robustness}
\end{equation}
upon establishing the abrupt rupture condition (\ref{eq:abrupt_condition_repeat}), i.e., that $h(x)$ takes its maximum at $x=S_{\textrm{min}}$. This follows immediately upon realizing that we have
\[
h(x) = \left \{ \begin{array}{ll}
x+ \bE{L},  & \textrm{for $0\leq x < \bE{S}$}     \\
 0,   &      \textrm{for $x \geq \bE{S}$}
\end{array} \right.
\]
under the distribution  (\ref{eq:dirac_osy}).
\myendpf


\bibliography{sample}


\section*{Acknowledgements}
This research was supported in part by the National Science Foundation through grant CCF \#1422165, and in part by the Department of Electrical and Computer Engineering at Carnegie Mellon University. O. Ya\u{g}an also acknowledges the Berkman Faculty Development Grant from Carnegie Mellon.

\section*{Author contributions statement}

Y.~Z. and O.~Y. have contributed equally to all parts of the paper. 

 \section*{Additional information}

\textbf{Competing financial interests.}
The authors declare no competing financial interests.






\newpage

\setcounter{equation}{0}
\setcounter{section}{0}

\renewcommand{\thesection}{\Alph{section}.}
\renewcommand{\theequation}{S.\arabic{equation}}

\begin{center}
{\LARGE{ SUPPLEMENTARY FILE}} 
\end{center}




\section{A proof of Claim \ref{claim:1}}
Let $x^{\star}$ denote the smallest solution of $x \geq g(x)$ 
and  $x^{\star\star}$ denote the smallest solution of  (\ref{eq:21}). Since 
$x \geq g(x)$ automatically gives (\ref{eq:21}), we always have $x^{\star\star} \leq x^{\star}$.
Here, our goal is to show that 
\begin{align}
\bP{S>x^{\star}} = \bP{S>x^{\star \star}}
\label{eq:claim_proof_to_show}
\end{align}
If $x^{\star}$ is also the smallest solution of (\ref{eq:21}), then the claim follows immediately by virtue of the fact that $x^{\star}=x^{\star \star}$.
Now, assume that there is a solution $x^{\star \star} < x^{\star}$ of (\ref{eq:21}). Then it must hold that $x^{\star \star}<g(x^{\star \star})$. From (\ref{eq:21}), this yields 
\[
1 = \mathbb{P}[S > g(x^{\star \star}) \mid S>x^{\star \star}] = \frac{\mathbb{P}[S>g(x^{\star \star}),S>x^{\star \star}]}{\mathbb{P}[S>x^{\star \star}]} 
=\frac{\mathbb{P}[S>g(x^{\star \star})]}{\mathbb{P}[S>x^{\star \star}]}
\]
Thus, we have 
\begin{align}
\mathbb{P}[S>x^{\star \star}] = \mathbb{P}[S>g(x^{\star \star})].
\label{eq:claim_proof_step1}
\end{align}

\begin{figure*}[!t]
	\includegraphics[width=0.5\textwidth]{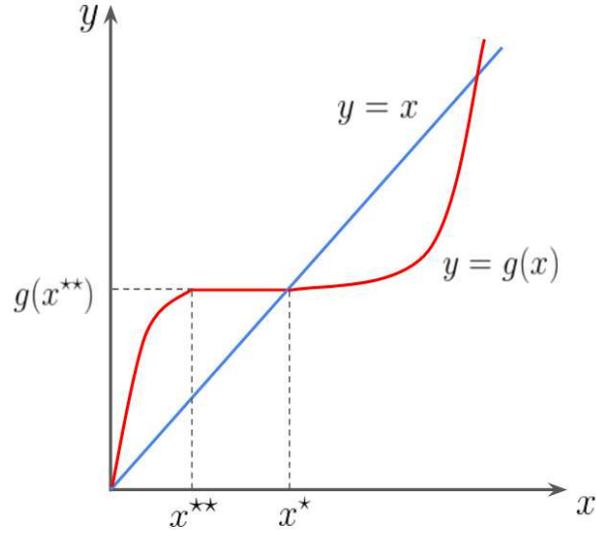}
    \centering
\caption{\sl \textbf{Illustration of function relations in Claim 1.} We define $x^{\star}$ as the first point where curves $y=x$ and $y=g(x)$ intersect. 
We show that if there exists $x^{\star \star}<x^{\star}$ satisfying (\ref{eq:21}), then $g(x)$ must be constant over $x^{\star \star} \leq x \leq x^{\star}$, yielding (\ref{eq:claim_proof_step2}).
}
    \label{fig:gx}
\end{figure*}

Key to the proof of Claim \ref{claim:1} is the observation that the function
$g(x)$ given at (\ref{eq:defn_g_x}) is monotone decreasing in $\bP{S>x}$. Put differently, (\ref{eq:claim_proof_step1}) implies that
$
g(x^{\star \star}) = g\left(g(x^{\star \star})\right)$,
meaning that $x=g(x^{\star \star})$ is a solution of $x \geq g(x)$. Since $x^{\star}$ is defined to be the smallest of all such solutions, this gives
\begin{equation}
x^{\star} \leq g(x^{\star \star}).
\label{eq:claim_proof_upper}
\end{equation}
On the other hand, the continuity assumption on the distribution of $S$ implies the continuity of $g(x)$. Hence, we have $x^{\star}=g(x^{\star})$. Recalling also that $g(x)$ is monotone increasing in $x$ and that $x^{\star \star} < x^{\star}$, we get
\begin{align}
g(x^{\star \star}) \leq g(x^{\star}) = x^{\star}.
\label{eq:claim_proof_lower}
\end{align}
Combining (\ref{eq:claim_proof_upper}) and (\ref{eq:claim_proof_lower}), we conclude that
\begin{align}
x^{\star} = g(x^{\star \star}).
\label{eq:claim_proof_step2}
\end{align}
Graphically, this means that the curve $y=g(x)$ is constant (and above the  line $y=x$) on the range  $[x^{\star \star}, x^{\star}]$  and it  intersects with the line $y=x$ at $x^{\star}$ (see figure \ref{fig:gx} for an illustration).

Combining (\ref{eq:claim_proof_step1}) and (\ref{eq:claim_proof_step2}), we get
\[
\mathbb{P}[S>x^{\star \star}] = \mathbb{P}[S>g(x^{\star \star})] = \bP{S> x^{\star}}
\]
which establishes the claim (\ref{eq:claim_proof_to_show}).
\myendpf

\section{A necessary condition for abrupt rupture}
Recall the condition (\ref{eq:abrupt_condition_repeat}) for an abrupt rupture to take place, namely the need for 
$h(x)$ to be maximized at $x=S_{\textrm{min}}$.
To explore this issue further, we now compute the derivative of $h(x)$. We have
\begin{align}
& \frac{d}{dx}(\mathbb{P}[S>x]( x+\mathbb{E}[L \mid S>x])) 
\nonumber\\
 &=\frac{d}{dx}\left( x\mathbb{P}[S>x]+\mathbb{E}[L \cdot  \mathbbm{1}[S>x]]\right) 
 \nonumber\\
 &= \mathbb{P}[S>x]+ x(-p_S(x))+\frac{d}{dx}\left(\int_{-\infty}^{\infty}\int_{-\infty}^{\infty} \ell \cdot \mathbbm{1}[s>x]p_{LS}(\ell,s)ds d\ell\right) 
 \nonumber\\
 &= \mathbb{P}[S>x] - xp_S(x)+\frac{d}{dx}\left(\int_{-\infty}^{\infty}\int_{x}^{\infty} \ell \cdot p_{LS}(\ell,s)dsd\ell\right) 
 \nonumber\\
 &= \mathbb{P}[S>x] -  xp_S(x)+\frac{d}{dx}\left(\int_{x}^{\infty}\int_{-\infty}^{\infty} \ell \cdot p_{LS}(\ell,s)d\ell ds\right) 
 \nonumber\\
 &= \mathbb{P}[S>x] -  xp_S(x) - \int_{-\infty}^{\infty} \ell \cdot p_{LS}(\ell,x)d\ell
  \nonumber\\
 &=\mathbb{P}[S>x] - xp_S(x) -\bE{L ~|~ S=x} p_S(x).
  \label{eq:27}
\end{align}

For $x<S_{\textrm{min}}$, we have $\mathbb{P}[S>x]=1$ and $\mathbb{E}[L \mid S>x]=\mathbb{E}[L]$, so that $ \frac{d}{dx} h(x)=1$;
this is already evident from (\ref{eq:h_x_till_S_min}). 
Then, for an abrupt rupture to take place, (\ref{eq:27}) should turn negative at $x = S_{\textrm{min}}$; i.e., it must hold that 
\[
 p_S(S_{\textrm{min}})(S_{\textrm{min}}  + \bE{L ~|~ S=S_{\textrm{min}}}) > 1.
\]
\myendpf


\end{document}